\documentclass[titlepage,12pt]{article}
\usepackage{amsmath}
\usepackage{amssymb}
\usepackage{supertabular}
\usepackage{yhmath}
\usepackage{url}

\newtheorem{theorem}{Theorem}
\newtheorem{lemma}{Lemma}
\newtheorem{corollary}{Corollary}

\newtheorem{proposition}{Proposition}

\def\vec#1{\mathchoice
{\mbox{\boldmath$\displaystyle#1$}} {\mbox{\boldmath$\textstyle#1$}}
{\mbox{\boldmath$\scriptstyle#1$}}
{\mbox{\boldmath$\scriptscriptstyle#1$}}}

\title{$m$-Sequences of Different Lengths with Four-Valued Cross Correlation}

\author{Tor Helleseth and Alexander Kholosha and Aina Johanssen\\
The Selmer Center,\\Department of Informatics, University of
Bergen\\PB 7800\\ N-5020 Bergen, Norway\\}

\begin{document}
\maketitle

\begin{quote}
{\bf Abstract.} Considered is the distribution of the cross
correlation between $m$-sequences of length $2^m-1$, where $m$ is
even, and $m$-sequences of shorter length $2^{m/2}-1$. The infinite
family of pairs of $m$-sequences with four-valued cross correlation
is constructed and the complete correlation distribution of this
family is determined.
\end{quote}

\begin{quote}
\end{quote}

\section{Introduction}
Let $\{a_t\}$ and $\{b_{t}\}$ be two binary sequences of length $p$.
The cross-correlation function between these two sequences at shift
$\tau$, where $0\leq\tau<p$, is defined by
\[C(\tau)=\sum_{t=0}^{p-1}(-1)^{a_{t}+b_{t+\tau}}\enspace.\]

A well studied problem is to find the cross-correlation function
between two binary $m$-sequences $\{s_t\}$ and $\{s_{dt}\}$ of the
same length $2^m-1$ that differ by a decimation $d$ such that
$\gcd(d,2^m-1)=1$. An overview of known results can be found in
Helleseth~\cite{He76}, Helleseth and Kumar~\cite{HeKu98} and
Dobbertin et. al.~\cite{DoFeHeRo06}.

Recently, Ness and Helleseth~\cite{NeHe06} studied the cross
correlation between any $m$-sequence $\{s_t\}$ of length $p=2^m-1$
and any $m$-sequences $\{u_{dt}\}$ of shorter length $2^{m/2}-1$,
where $m$ is even and $\gcd(d,2^{m/2}-1)=1$. For convenience,
$\{u_t\}$ is selected to be the $m$-sequence used in the small
Kasami sequence family. The only known families of $m$-sequences of
these periods giving a two-valued cross correlation are related to
the Kasami sequences~\cite{Ka66} and are obtained taking $d=1$.
Further, families with three-valued cross correlation have been
constructed by Ness and Helleseth in \cite{NeHe06} and
\cite{NeHe06_1}. These results were generalized by Helleseth,
Kholosha and Ness~\cite{HeKhNe07} who covered all known cases of
three-valued cross correlation and conjectured that these were the
only existing.

In this paper, we consider pairs of sequences with a four-valued
cross correlation. The first family with such a property was
described in \cite{NeHe07}. We completed a full search for all
values of $m\leq 32$ and revealed a few examples that did not fit
into the known family. Most of the cases with four-valued cross
correlation occur for $m=2nk$ with $n>2$ odd and the decimation
\[d=\frac{2^{nk}+1}{2^k+1}\quad\mbox{with}\quad k>1\enspace.\]
The main result of this paper is finding the distribution for this
four-valued cross correlation. Note that the family found in
\cite{NeHe07} corresponds to the latter decimation when setting
$n=3$.

In Section~\ref{sec:Prel}, we present preliminaries needed to prove
our main results. In Section~\ref{sec:Aff}, we give the distribution
of the number of zeros of a particular affine polynomial $A_a(x)$.
Section~\ref{sec:Lin} provides the distribution of the number of
zeros of a special linearized polynomial $L_a(z)$. The zeros of
these two polynomials are useful when obtaining the
cross-correlation values. Section~\ref{sec:4VDist} determines the
cross-correlation distribution of our four-valued family.

\section{Preliminaries}
 \label{sec:Prel}
Let $\mathrm{GF}(q)$ denote a finite field with $q$ elements and let
$\mathrm{GF}(q)^*=\mathrm{GF}(q)\setminus\{0\}$. The finite field
$\mathrm{GF}(q^l)$ is a subfield of $\mathrm{GF}(q^m)$ if and only
if $l$ divides $m$. The trace mapping from $\mathrm{GF}(q^m)$ to the
subfield $\mathrm{GF}(q^l)$ is defined by
\[{\rm Tr}_l^m(x)=\sum_{i=0}^{m/l-1}x^{q^{li}}\enspace.\]
In the case when $l=1$, we use the notation ${\rm Tr}_m(x)$ instead
of ${\rm Tr}_1^m(x)$. The norm ${\rm N}_l^m(x)$ of
$x\in\mathrm{GF}(q^m)$ over the subfield $\mathrm{GF}(q^l)$ is
defined by
\[{\rm N}_l^m(x)=\prod_{i=0}^{m/l-1}x^{q^{li}}\enspace.\]

Let $m$ be even and $\alpha$ be an element of order $p=2^m-1$ in
$\mathrm{GF}(2^m)$. Then the $m$-sequence $\{s_t\}$ of length
$p=2^m-1$ can be written in terms of the trace mapping as
\[s_t={\rm Tr}_m(\alpha^t)\enspace.\]
Let $\beta=\alpha^{2^{m/2}+1}$ be an element of order $2^{m/2}-1$.
The sequence $\{u_t\}$ of length $2^{m/2}-1$ (which is used in the
construction of the well known Kasami family) is defined by
\[u_t={\rm Tr}_{m/2}(\beta^t)\enspace.\]

In this paper, we consider the cross correlation between the
$m$-sequences $\{s_t\}$ and $\{v_t\}=\{u_{dt}\}$ at shift $\tau$
defined by
\begin{equation}
 \label{eq:Cd}
C_d(\tau)=\sum_{t=0}^{p-1}(-1)^{s_t+v_{t+\tau}}\enspace,
\end{equation}
where $\gcd(d,2^{m/2}-1)=1$ and $\tau=0,1,\dots,2^{m/2}-2$. Using
the trace representation, Ness and Helleseth~\cite{NeHe06} showed
that the set of values of $C_d(\tau)+1$ for
$\tau=0,1,\dots,2^{m/2}-2$ is equal to the set of values of
\begin{equation}
 \label{eq:S}
S(a)=\sum_{x\in\mathrm{GF}(2^m)}(-1)^{{\rm Tr}_m(ax)+{\rm
Tr}_{m/2}(x^{d(2^{m/2}+1)})}
\end{equation}
when $a\in\mathrm{GF}(2^{m/2})^*$.

The main result of this paper is formulated in the following
corollary that gives a four-valued cross-correlation function
between new pairs of $m$-sequences of different lengths.

\begin{corollary}
 \label{co:main}
Let $m=2nk$ and $d=\frac{2^{nk}+1}{2^k+1}$, where $n>2$ is odd and
$k>1$. Then the cross-correlation function $C_d(\tau)$ has the
following distribution:
\[\begin{array}{llll}
-1-2^{(n+1)k}&\mbox{occurs}&\frac{2^{(n-1)k}-1}{2^{2k}-1}&\mbox{times}\ ,\\
-1-2^{nk}&\mbox{occurs}&\frac{(2^{nk}-1)(2^{k-1}-1)}{2^k-1}&\mbox{times}\ ,\\
-1&\mbox{occurs}&2^{(n-1)k}-1&\mbox{times}\ ,\\
-1+2^{nk}&\mbox{occurs}&\frac{(2^{nk}+1)2^{k-1}}{2^k+1}&\mbox{times}\enspace.
\end{array}\]
\end{corollary}

The result will be proved in a series of lemmas and propositions.
The outline of the proof is as follows. Determining the set of
values of $C_d(\tau)+1$ for $\tau=0,1,\dots,2^{nk}-2$ is equivalent
to finding the set of values of $S(a)$ in (\ref{eq:S}) for
$a\in\mathrm{GF}(2^{nk})^*$. Furthermore, we show that
\[S(a)=\frac{1}{2^k+1}\sum_{i=0}^{2^k}S_i(a)\enspace,\]
where $S_i(a)$ are defined by
 \setlength{\arraycolsep}{0.14em}
\begin{eqnarray*}
S_j(a)&=&\sum_{y\in\mathrm{GF}(2^m)}(-1)^{{\rm Tr}_m(r^j
ay^{2^k+1}) +{\rm Tr}_{nk}(y^{2^{nk}+1})}\quad\mbox{and}\\
S_{2^k+1-j}(a)&=&\sum_{y\in\mathrm{GF}(2^m)}(-1)^{{\rm
Tr}_m(r^{-j}ay^{2^k+1}) +{\rm Tr}_{nk}(y^{2^{nk}+1})}
\end{eqnarray*}
 \setlength{\arraycolsep}{5pt}\noindent
for $j=0,1,\dots,2^{k-1}$ and with $r=\alpha^{(2^{nk}-1)2^{k-1}}$.

We determine $S_0(a)$ exactly in Corollary~\ref{cor:S0} and find
$S_i(a)^2$ in Lemma~\ref{le:SiSquare}. Since $S(a)$ is an integer,
we can resolve the sign ambiguity of all $S_i(a)$ for
$i=1,2,\ldots,2^k$. In order to determine $S_0(a)$, we need to
consider zeros in $\mathrm{GF}(2^{nk})$ of the affine polynomial
\[A_a(x)=a^{2^k} x^{2^{2k}}+x^{2^k}+ax+c\enspace,\]
where $c\in\mathrm{GF}(2^k)$ and ${\rm Tr}_k(c)=1$. To determine
$S_i(a)^2$ for $1=1,2,\dots,2^k$, we need to consider zeros in
$\mathrm{GF}(2^{2nk})$ of the linearized polynomial
\[L_a(z)=z^{2^{(n+1)k}}+r^{2^k}a^{2^k}z^{2^{2k}}+raz\enspace,\]
where $n$ is odd, $a\in\mathrm{GF}(2^{nk})$ and
$r\in\mathrm{GF}(2^{2nk})^*$ with $r^{2^{nk}+1}=1$ but
$r^{\frac{2^{nk}+1}{2^k+1}}\neq 1$.

When finding the complete cross-correlation distribution, we make
use of the following lemma from \cite{NeHe06}.

\begin{lemma}[\cite{NeHe06}]
 \label{le:pow1}
For any decimation $d$ with $\gcd(d,2^{nk}-1)=1$ the sum of the
cross-correlation values defined in (\ref{eq:Cd}) for $m=2nk$ is
equal to
\[\sum_{\tau=0}^{2^{nk}-2} C_d(\tau)=1\enspace.\]
\end{lemma}

\section{The Affine Polynomial $A_a(x)$}
 \label{sec:Aff}
In this section, we consider zeros in $\mathrm{GF}(2^{nk})$, with
$n>2$, of the affine polynomial
\begin{equation}
 \label{eq:A}
A_a(x)=a^{2^k}x^{2^{2k}}+x^{2^k}+ax+c\enspace,
\end{equation}
where $a\in\mathrm{GF}(2^{nk})$ and $c\in\mathrm{GF}(2^k)$. Some
additional conditions on the parameters will be imposed later. The
distribution of zeros in $\mathrm{GF}(2^{nk})$ of (\ref{eq:A}) will
determine to a large extent the distribution of our
cross-correlation function. It is clear that $A_a(x)$ does not have
multiple roots if $a\neq 0$.

We introduce a particular sequence of polynomials over
$\mathrm{GF}(2^{nk})$ that will play a crucial role when finding
zeros of (\ref{eq:A}). First, for any $v\in\mathrm{GF}(2^{nk})$
denote $v_i=v^{2^{ik}}$ for $i=0,\dots,n-1$ so $A_a(x)=a_1
x_2+x_1+a_0 x_0+c$. Let
 \setlength{\arraycolsep}{0.14em}
\begin{eqnarray}
 \label{eq:dB1}
\nonumber B_1(x)&=&1\ ,\\
\nonumber B_2(x)&=&1\ ,\\
B_{i+2}(x)&=&B_{i+1}(x)+x_i B_i(x)\quad\mbox{for}\quad 1\leq i\leq
n-1\enspace.
\end{eqnarray}
 \setlength{\arraycolsep}{5pt}\noindent

Observe the following recursive identity that can be seen as an
equivalent definition of $B_i(x)$
\begin{equation}
 \label{eq:dB2}
B_{i+2}(x)=B_{i+1}^{2^k}(x)+x_1 B_i^{2^{2k}}(x)\quad\mbox{for}\quad
1\leq i\leq n-1\enspace.
\end{equation}
We prove it using induction on $i$. For $i=1$ and $i=2$ this fact is
easily checked taking the definition. Assuming this identity holds
for $i<t$ we get for $i=t>2$
 \setlength{\arraycolsep}{0.14em}
\begin{eqnarray*}
B_{t+2}(x)&\stackrel{(\ref{eq:dB1})}{=}&B_{t+1}(x)+x_t B_t(x)\\
&=&B_t^{2^k}(x)+x_1 B_{t-1}^{2^{2k}}(x)+x_t B_{t-1}^{2^k}(x)+x_t x_1 B_{t-2}^{2^{2k}}(x)\\
&=&(B_t(x)+x_{t-1} B_{t-1}(x))^{2^k}+x_1(B_{t-1}(x)+x_{t-2} B_{t-2}(x))^{2^{2k}}\\
&\stackrel{(\ref{eq:dB1})}{=}&B_{t+1}^{2^k}(x)+x_1
B_t(x)^{2^{2k}}\enspace.
\end{eqnarray*}
 \setlength{\arraycolsep}{5pt}\noindent

We also define polynomials $Z_n(x)$ over $\mathrm{GF}(2^{nk})$ as
\begin{equation}
 \label{eq:Z}
Z_n(x)=B_{n+1}(x)+x B_{n-1}^{2^k}(x)\enspace.
\end{equation}
The following lemma describes zeros of $B_n(x)$ and $Z_n(x)$ in
$\mathrm{GF}(2^{nk})$.

\begin{lemma}
 \label{le:ZeB}
For any $v\in\mathrm{GF}(2^{nk})\setminus\mathrm{GF}(2^k)$ let
\begin{equation}
 \label{eq:V}
V=\frac{v_0^{2^{2k}+1}}{(v_0+v_1)^{2^k+1}}\enspace.
\end{equation}
Then for $n>1$
\[B_n(V)=\frac{{\rm
Tr}^{nk}_k(v_0)}{(v_1+v_2)}\prod_{j=2}^{n-1}\left(\frac{v_0}{v_0+v_1}\right)^{2^{jk}}\enspace.\]
If $n>1$ is odd (resp. $n>2$ is even) then the total number of
distinct zeros of $B_n(x)$ in $\mathrm{GF}(2^{nk})$ is equal to
$\frac{2^{(n-1)k}-1}{2^{2k}-1}$ (resp.
$\frac{2^{(n-1)k}-2^k}{2^{2k}-1}$). Moreover, polynomial $B_n(x)$
splits in $\mathrm{GF}(2^{nk})$, all its zeros have the form of
(\ref{eq:V}) with ${\rm Tr}^{nk}_k(v_0)=0$ and occur with
multiplicity $2^k$.
\end{lemma}

{\bf Proof.} First, note that
$v\in\mathrm{GF}(2^{nk})\setminus\mathrm{GF}(2^k)$ if and only if
$v_0\neq v_1$ which guarantees that the denominator in (\ref{eq:V})
and in the above identity for $B_n(V)$ is not zero. Now, using
induction on $i$ we prove that
\begin{equation}
 \label{eq:BiV}
B_i(V)=\frac{\sum_{j=1}^i
v_j}{(v_1+v_2)}\prod_{j=2}^{i-1}\left(\frac{v_0}{v_0+v_1}\right)^{2^{jk}}
\end{equation}
for $2\leq i\leq n+1$. For $i=2$ and $i=3$ this identity is easily
checked using the definition (\ref{eq:dB1}) of $B_i(x)$ (for $i=2$,
we assume the product over the empty set to be equal to $1$).
Assuming this identity holds for $i<t$ we get for $i=t>3$
 \setlength{\arraycolsep}{0.14em}
\begin{eqnarray*}
B_t(V)&\stackrel{(\ref{eq:dB1})}{=}&B_{t-1}(V)+V_{t-2} B_{t-2}(V)\\
&=&\frac{\sum_{j=1}^{t-1}v_j}{(v_1+v_2)}\prod_{j=2}^{t-2}\left(\frac{v_0}{v_0+v_1}\right)^{2^{jk}}+
\frac{v_{t-2}^{2^{2k}+1}\sum_{j=1}^{t-2}v_j}{(v_{t-2}+v_{t-1})^{2^k+1}(v_1+v_2)}
\prod_{j=2}^{t-3}\left(\frac{v_0}{v_0+v_1}\right)^{2^{jk}}\\
&=&\frac{\left((v_{t-1}+v_t)\sum_{j=1}^{t-1}v_j+v_t\sum_{j=1}^{t-2}v_j\right)\prod_{j=2}^{t-2}v_0^{2^{jk}}}
{(v_1+v_2)\prod_{j=2}^{t-1}(v_0+v_1)^{2^{jk}}}\\
&=&\frac{\sum_{j=1}^t
v_j}{(v_1+v_2)}\prod_{j=2}^{t-1}\left(\frac{v_0}{v_0+v_1}\right)^{2^{jk}}\enspace.
\end{eqnarray*}
 \setlength{\arraycolsep}{5pt}\noindent
It remains to note that for $i=n$, in $\mathrm{GF}(2^{nk})$ we have
$\sum_{j=1}^n v_j={\rm Tr}^{nk}_k(v_0)$.

Obviously, $B_n(V)=0$ if and only if ${\rm Tr}^{nk}_k(v_0)=0$ which
is equivalent to $v_0=u+u^{2^k}$ for some
$u\in\mathrm{GF}(2^{nk})\setminus\mathrm{GF}(2^{2k})$ (since
$v_0\in\mathrm{GF}(2^k)$ if and only if the corresponding
$u\in\mathrm{GF}(2^{2k})$). It follows from the proof of
Proposition~\ref{pr:l22kZero} that the mapping from
$u\in\mathrm{GF}(2^{nk})\setminus\mathrm{GF}(2^{2k})$ via
$v_0=u+u^{2^k}$ to $V\in\mathrm{GF}(2^{nk})^*$ defined by
(\ref{eq:V}) is $(2^{3k}-2^k)$-to-$1$. Therefore, we have found
$\frac{|\mathrm{GF}(2^{nk})\setminus\mathrm{GF}(2^{2k})|}{2^{3k}-2^k}$
distinct zeros of $B_n(x)$ in $\mathrm{GF}(2^{nk})$ and if $n$ is
odd (resp. $n$ is even) then this number is equal to
$\frac{2^{(n-1)k}-1}{2^{2k}-1}$ (resp.
$\frac{2^{(n-1)k}-2^k}{2^{2k}-1}$).

It is easy to check by induction that if $i$ is odd (resp. $i$ is
even) then the algebraic degree of polynomials $B_i(x)$ is equal to
$\frac{2^{ik}-2^k}{2^{2k}-1}$ (resp.
$\frac{2^{ik}-2^{2k}}{2^{2k}-1}$) since
\[\deg B_{i+2}(x)=\max\{\deg B_{i+1}(x),2^{ik}+\deg B_i(x)\}=2^{ik}+\deg B_i(x)\enspace.\]
Further, if we define the sequence of polynomials $B'_i(x)$ for
$i=1,\dots,n$ with $B'_1(x)=B'_2(x)=1$ and
$B'_{i+2}(x)=B'_{i+1}(x)+x_{i-1}B'_i(x)$ then $B_i(x)=B'_i(x)^{2^k}$
for $i=1,\dots,n$. Therefore, all zeros of $B_n(x)$ having the form
of (\ref{eq:V}) with ${\rm Tr}^{nk}_k(v_0)=0$ have multiplicity at
least $2^k$. Finally, note that the number of these zeros multiplied
by $2^k$ is equal to the degree of $B_n(x)$.
\hspace*{\fill}${\Box}$

\begin{corollary}
 \label{co:ZeZ}
For any $n>1$, polynomial $Z_n(x)$ splits in $\mathrm{GF}(2^{nk})$
with all its zeros having the form of (\ref{eq:V}) and without
multiple roots. If $n$ is odd (resp. $n$ is even) then the total
number of zeros of $Z_n(x)$ in $\mathrm{GF}(2^{nk})$ is equal to
$\frac{2^{(n+1)k}-2^{2k}}{2^{2k}-1}$ (resp.
$\frac{2^{(n+1)k}-2^k}{2^{2k}-1}$).
\end{corollary}

{\bf Proof.} Using (\ref{eq:BiV}), it can be verified directly that
$B_{n+1}(V)=V B_{n-1}^{2^k}(V)$ for any $V\in\mathrm{GF}(2^{nk})$
having the form of (\ref{eq:V}) (the case $n=2$ is easily checked
having the definition of $B_i(x)$). Also, using the fact from the
latest proof, we conclude that $\deg Z_n(x)=\deg B_{n+1}(x)$ and is
equal to $\frac{2^{(n+1)k}-2^{2k}}{2^{2k}-1}$ (resp.
$\frac{2^{(n+1)k}-2^k}{2^{2k}-1}$) if $n$ is odd (resp. $n$ is
even). Denote $S=\{x\in\mathrm{GF}(2^{nk})\setminus\mathrm{GF}(2^k)\
|\ {\rm Tr}^{nk}_k(x)\neq 0\}$. It follows from the proof of
Proposition~\ref{pr:A2kZero} that the mapping from $v\in S$ to
$V\in\mathrm{GF}(2^{nk})^*$ defined by (\ref{eq:V}) is
$(2^k-1)$-to-$1$. Recalling the corresponding fact from the latest
proof, we conclude that the total number of distinct values of $V$
obtained by (\ref{eq:V}) is equal to
$\frac{|\mathrm{GF}(2^{nk})\setminus\mathrm{GF}(2^{2k})|}{2^{3k}-2^k}+\frac{|S|}{2^k-1}$
being identical to the degree of $Z_n(x)$. Note that two different
values of $v\in\mathrm{GF}(2^{nd})\setminus\mathrm{GF}(2^d)$ with
zero and nonzero trace in $\mathrm{GF}(2^d)$ can not map to the same
value $V$ using (\ref{eq:V}) since $C_n(V)=0$ if and only if the
trace of the corresponding $v$ is also equal zero.
\hspace*{\fill}${\Box}$

\begin{corollary}
 \label{co:ZeTr}
For any $V\in\mathrm{GF}(2^{nk})$ having the form of (\ref{eq:V})
with $n>2$ and ${\rm Tr}^{nk}_k(v_0)\neq 0$ we have
\[{\rm
Tr}_k^{nk}\left(\frac{B_{n-1}^{2^k}(V)}{B_n^{2^k+1}(V)}\right)=0={\rm
Tr}_k^{nk}\left(\frac{B_{n-1}^{2^k}(V)B_{n+1}(V)}{B_n^{2^{k}+1}(V)}\right)\enspace,\]
where the second identity holds if and only if $n$ is odd.
\end{corollary}

{\bf Proof.} Using (\ref{eq:BiV}), it can be verified directly that
 \setlength{\arraycolsep}{0.14em}
\begin{eqnarray*}
\frac{B_{n-1}^{2^k}(V)}{B_n^{2^k+1}(V)}&=&{\rm
N}_k^{nk}\left(1+\frac{v_1}{v_0}\right)\frac{v_1\sum_{j=2}^n
v_j}{{\rm Tr}_k^{nk}(v_0)^2}\quad\quad\mbox{and}\\
\frac{B_{n-1}^{2^k}(V)B_{n+1}(V)}{B_n^{2^{k}+1}(V)}&=&
\frac{\left(v_1+{\rm Tr}_k^{nk}(v_0)\right)\sum_{j=2}^n v_j}{{\rm
Tr}_k^{nk}(v_0)^2}
\end{eqnarray*}
 \setlength{\arraycolsep}{5pt}\noindent
for any $V\in\mathrm{GF}(2^{nk})$ having the form of (\ref{eq:V})
and $n>2$. Now note that
\[{\rm Tr}_k^{nk}\Big(v_1 \sum_{j=2}^n v_j\Big)={\rm Tr}_k^{nk}\left(v_1
{\rm Tr}_k^{nk}(v_0)+v_1^2\right)={\rm Tr}_k^{nk}(v_0)^2+{\rm
Tr}_k^{nk}(v_0^2)=0\] and thus,
\[{\rm Tr}_k^{nk}\Big(\left(v_1+{\rm Tr}_k^{nk}(v_0)\right)\sum_{j=2}^n v_j\Big)=
{\rm Tr}_k^{nk}(v_0){\rm Tr}_k^{nk}\left(({\rm
Tr}_k^{nk}(v_0)+v_1)\right)=0\] if $n$ is odd (and equal to ${\rm
Tr}_k^{nk}(v_0)^2\neq 0$ if $n$ is even).
\hspace*{\fill}${\Box}$

Polynomials $B_i(x)$ can be interpreted as the determinant of
three-diagonal symmetric matrices (note a comprehensive study of
these matrices in \cite{IlKu85}). Indeed, for $j\leq i$ let
$\Delta_x(j,i)$ denote the determinant of matrix $D_x$ of size
$i-j+2$ that contains ones on the main diagonal and with
$D_x(t,t+1)=D_x(t+1,t)=x_{j+t-1}$ for $t=1,\dots,i-j+1$, where the
indices of $x_i$ are reduced modulo $n$. Expanding the determinant
of $D_x$ by minors along the last row we obtain
\begin{equation}
 \label{eq:RecDelta}
\Delta_x(j,i)=\Delta_x(j,i-1)+x_i^2\Delta_x(j,i-2)
\end{equation}
assuming $\Delta_x(j,i)=1$ if $i-j\in\{-2,-1\}$. Comparing the
latter recursive identity with (\ref{eq:dB1}) it is easy to see that
\begin{equation}
 \label{eq:Delta}
\Delta_x(1,i)=B_{i+2}^2(x)\enspace.
\end{equation}
Moreover, from the definition of the determinant it also follows
that
\begin{equation}
 \label{eq:Delta2k}
\Delta_x(1,i)^{2^{tk}}=\Delta_x(1+t,i+t)\quad\mbox{for}\quad 0\leq
t\leq n-1\enspace.
\end{equation}

We will need the following result that can be obtained combining
Theorems~5.6 and 6.4 in \cite{Bl04}.

\begin{theorem}[\cite{Bl04}]
 \label{th:Bl}
Take polynomials over $\mathrm{GF}(2^{nk})$
\[f(x)=x^{2^k+1}+b^2 x+b^2\quad{\mbox and}\quad
g(x)=b^{-1}f(bx^{2^k-1})=b^{2^k}x^{2^{2k}-1}+b^2 x^{2^k-1}+b\] with
$b\neq 0$. Then exactly one of the following holds
\renewcommand{\theenumi}{\roman{enumi}}
\renewcommand{\labelenumi}{(\theenumi)}
\begin{enumerate}
\item\label{it:1} $f(x)$ has none or two zeros in $\mathrm{GF}(2^{nk})$ and
g(x) has none zeros in $\mathrm{GF}(2^{nk})$;

\item\label{it:2} $f(x)$ has one zero in $\mathrm{GF}(2^{nk})$ and g(x) has
$2^k-1$ zeros in $\mathrm{GF}(2^{nk})$;

\item\label{it:3} $f(x)$ has $2^k+1$ zeros in $\mathrm{GF}(2^{nk})$ and
g(x) has $2^{2k}-1$ zeros in $\mathrm{GF}(2^{nk})$.
\end{enumerate}
Let $N_i$ denote the number of $b\in\mathrm{GF}(2^{nk})^*$ such that
$g(x)=0$ has exactly $i$ roots in $\mathrm{GF}(2^{nk})$. Then the
following distribution holds for $n$ odd (resp. $n$ even)
\[\begin{array}{llll}
N_0&=&\frac{2^{(n+2)k}-2^{(n+1)k}-2^{nk}+1}{2^{2k}-1}&\quad
(\mbox{resp.}\
\frac{2^{(n+2)k}-2^{(n+1)k}-2^{nk}-2^{2k}+2^k+1}{2^{2k}-1})\ ,\\
N_{2^k-1}&=&2^{(n-1)k}-1&\quad (\mbox{resp.}\ 2^{(n-1)k})\ ,\\
N_{2^{2k}-1}&=&\frac{2^{(n-1)k}-1}{2^{2k}-1}&\quad (\mbox{resp.}\
\frac{2^{(n-1)k}-2^k}{2^{2k}-1})\enspace.
\end{array}\]
\end{theorem}

Let
\[M_i=\{a\;|\;a\neq 0, A_a(x)\ \mbox{has exactly $i$ zeros in}\ \mathrm{GF}(2^{nk})\}\enspace.\]
Obviously, either $A_a(x)$ has no zeros in $\mathrm{GF}(2^{nk})$ or
it has exactly the same number of zeros as its linearized
homogeneous part that is $l_a(x)=a_1 x^{2^{2k}}+x^{2^k}+a_0 x$. The
zeros in $\mathrm{GF}(2^{nk})$ of $l_a(x)$ form a vector subspace
over $\mathrm{GF}(2^k)$. In the following propositions, we prove
that $A_a(x)$ always has a zero in $\mathrm{GF}(2^{nk})$ so $A_a(x)$
and $l_a(x)$ have the same number of zeros in $\mathrm{GF}(2^{nk})$
that can be equal to $1,2^k$ or $2^{2k}$. Assume $a\neq 0$, then
dividing $l_a(x)$ by $a_0 a_1 x$ (we remove one zero $x=0$) and then
substituting $x$ with $a_0^{-1}x$ leads to $a_1^{-2^k}
x^{2^{2k}-1}+a_1^{-2} x^{2^k-1}+a_1^{-1}$ which has the form of
polynomial $g(x)$ from Theorem~\ref{th:Bl} taking $b=a_1^{-1}$ (note
a $1$-to-$1$ correspondence between $a$ and $b$). Thus,
$|M_i|=N_{i-1}$ for $i\in\{1,2^k,2^{2k}\}$.

\begin{proposition}
 \label{pr:A1Zero}
For any $a\in\mathrm{GF}(2^{nk})^*$, polynomial $A_a(x)$ has exactly
one zero in $\mathrm{GF}(2^{nk})$ if and only if $Z_n(a)\neq 0$.
Moreover, this zero is equal to $\mathcal{V}_a=c B_n(a)/Z_n(a)$ and
${\rm Tr}_{nk}(\mathcal{V}_a)={\rm Tr}_k(nc)$. Also if $n$ is odd
(resp. $n$ is even) then
\[|M_1|=\frac{2^{(n+2)k}-2^{(n+1)k}-2^{nk}+1}{2^{2k}-1}\ (\mbox{resp.}\
\frac{2^{(n+2)k}-2^{(n+1)k}-2^{nk}-2^{2k}+2^k+1}{2^{2k}-1})\ .\]
\end{proposition}

{\bf Proof.} We start with proving that $c B_n(a)/Z_n(a)$ indeed is
a zero of $A_a(x)$ if $Z_n(a)\neq 0$. First, for any
$v\in\mathrm{GF}(2^{nk})$, using both recursive definitions of
$B_n(x)$
 \setlength{\arraycolsep}{0.14em}
\begin{eqnarray*}
Z_n^{2^k}(v)&\stackrel{(\ref{eq:Z})}{=}&B_{n+1}^{2^k}(v)+v_1 B_{n-1}^{2^{2k}}(v)\\
&\stackrel{(\ref{eq:dB1})}{=}&B_n^{2^k}(v)+v_0 B_{n-1}^{2^k}(v)+v_1 B_{n-1}^{2^{2k}}(v)\\
&\stackrel{(\ref{eq:dB2})}{=}&B_{n+1}(v)+v_0 B_{n-1}^{2^k}(v)\\
&\stackrel{(\ref{eq:Z})}{=}&Z_n(v)
\end{eqnarray*}
 \setlength{\arraycolsep}{5pt}\noindent
and thus, $Z_n(v)\in\mathrm{GF}(2^k)$. Therefore,
 \setlength{\arraycolsep}{0.14em}
\begin{eqnarray}
 \label{eq:A(B)}
A_a(\mathcal{V}_a)
&=&\frac{c}{Z_n(a)}\left(a_1 B_n^{2^{2k}}(a)+B_n^{2^k}(a)+a_0 B_n(a)+Z_n(a)\right)\\
\nonumber&\stackrel{(\ref{eq:dB1})}{=}&\frac{c}{Z_n(a)}\left(a_1
B_{n-1}^{2^{2k}}(a)+a_1 a_0
B_{n-2}^{2^{2k}}(a)+B_n^{2^k}(a)+a_0 B_n(a)+Z_n(a)\right)\\
\nonumber&\stackrel{(\ref{eq:dB2})}{=}&\frac{c}{Z_n(a)}\left(B_{n+1}(a)+a_0
B_{n-1}^{2^k}(a)+Z_n(a)\right)=0\enspace.
\end{eqnarray}
 \setlength{\arraycolsep}{5pt}\noindent

Now we show that in our case $\mathcal{V}_a$ is the only zero of
$A_a(x)$. Taking equation $A_a(x)=0$ and all its $2^{ik}$ powers we
obtain $n$ equations
\[A_a^{2^{ik}}(x)=a_{i+1} x_{i+2}+x_{i+1}+a_i x_i+c=0\quad\mbox{for}\quad
i=0,\dots,n-1\enspace,\] where all indices are calculated modulo
$n$. If $x_i$ $(i=0,\dots,n-1)$ are considered as independent
variables then the obtained system of $n$ linear equations with $n$
unknowns has the following matrix with the antidiagonal structure
\begin{equation}
 \label{eq:matr}
\left(\begin{array}{cccccc}
0&0&\cdots&a_1&1&a_0\\
0&&\adots&1&a_1&0\\
\vdots&\adots&\adots&\adots&\adots&\vdots\\
a_{n-2}&1&\adots&&&0\\
1&a_{n-2}&\adots&&0&a_{n-1}\\
a_{n-1}&0&\cdots&0&a_0&1
\end{array}\right)\enspace.
\end{equation}
Let the columns of (\ref{eq:matr}) be numbered from $1$ to $n$.
Permuting the columns in (\ref{eq:matr}) (reorder them as
$n-1,n-2,\dots,1,n$) we obtain the symmetric three-diagonal cyclic
matrix $\mathcal{M}_n$ containing ones on the main diagonal, with
$\mathcal{M}_n(i,i+1)=\mathcal{M}_n(i+1,i)=a_i$ for $i=1,\dots,n-1$
and with corner elements
$\mathcal{M}_n(1,n)=\mathcal{M}_n(n,1)=a_0$. If
$\vec{x}=(x_1,\dots,x_{n-1},x_0)^{\rm T}$ and
$\vec{c}=(c,\dots,c)^{\rm T}$ then the system has the following
matrix representation
\begin{equation}
 \label{eq:syst}
\mathcal{M}_n\vec{x}=\vec{c}\enspace.
\end{equation}
The determinant of (\ref{eq:matr}) is equal to the determinant of
$\mathcal{M}_n$ and can be computed expanding the latter by minors
along the last row. Doing this it is easy to see that
 \setlength{\arraycolsep}{0.14em}
\begin{eqnarray*}
\det\mathcal{M}_n&=&\Delta_a(1,n-2)+a_{n-1}(a_{n-1}\Delta_a(1,n-3)+a_0\dots a_{n-2})\\
&&\quad\quad\quad\quad\quad\ \,+\;a_0(a_0\Delta_a(2,n-2)+a_1\dots a_{n-1})\\
&\stackrel{(\ref{eq:Delta},\ref{eq:Delta2k})}{=}&B_n^2(a)+a_{n-1}^2 B_{n-1}^2(a)+(a_0 B_{n-1}^{2^k}(a))^2\\
&\stackrel{(\ref{eq:dB1})}{=}&B_{n+1}^2(a)+(a_0 B_{n-1}^{2^k}(a))^2\\
&\stackrel{(\ref{eq:Z})}{=}&Z_n^2(a)\enspace.
\end{eqnarray*}
 \setlength{\arraycolsep}{5pt}\noindent
Thus, if $Z_n(a)\neq 0$ then (\ref{eq:syst}) has exactly one
solution. Now note that every $v\in\mathrm{GF}(2^{nk})$ with
$A_a(v)=0$ provides a solution to the system given by
$v_i=v^{2^{ik}}$ for $i=0,\dots,n-1$. Therefore, if $Z_n(a)\neq 0$
then $A_a(x)$ has at most one zero.

Using Corollary~\ref{co:ZeZ}, we can obtain the number of
$a\in\mathrm{GF}(2^{nk})^*$ such that $Z_n(a)\neq 0$ (note that
$Z_n(0)=1$). Observe that this number is identical to $N_0$ from
Theorem~\ref{th:Bl} that is equal to the number of
$a\in\mathrm{GF}(2^{nk})^*$ such that $l_a(x)=0$ has exactly one
root in $\mathrm{GF}(2^{nk})$ (see explanations following
Theorem~\ref{th:Bl}). Therefore, if $A_a(x)$ has exactly one zero in
$\mathrm{GF}(2^{nk})$ then its homogeneous part $l_a(x)$ has the
same number of zeros and so $a$ is necessarily such that $Z_n(a)\neq
0$.

Finally, to prove the trace identity for $\mathcal{V}_a$ first note
that for any $v\in\mathrm{GF}(2^{nk})$
 \setlength{\arraycolsep}{0.14em}
\begin{eqnarray}
 \label{eq:TrBZ}
{\rm Tr}_k^{nk}(B_n(v)+Z_n(v))&\stackrel{(\ref{eq:Z})}{=}&{\rm Tr}_k^{nk}\left(B_n(v)+B_{n+1}(v)+v_0 B_{n-1}^{2^k}(v)\right)\\
\nonumber&\stackrel{(\ref{eq:dB1})}{=}&{\rm Tr}_k^{nk}\left(B_n(v)+B_n(v)+v_{n-1} B_{n-1}(v)+v_0 B_{n-1}^{2^k}(v)\right)\\
\nonumber&=&{\rm Tr}_k^{nk}\left(v_{n-1} B_{n-1}(v)+(v_{n-1}
B_{n-1}(v))^{2^k}\right)=0\enspace.
\end{eqnarray}
 \setlength{\arraycolsep}{5pt}\noindent
Therefore, since $c$ and $Z_n(v)$ are both in $\mathrm{GF}(2^k)$,
then
 \setlength{\arraycolsep}{0.14em}
\begin{eqnarray*}
{\rm Tr}_{nk}(\mathcal{V}_v)&=&
{\rm Tr}_{nk}\left(c+c\,\frac{B_n(v)+Z_n(v)}{Z_n(v)}\right)\\
&=&{\rm Tr}_k(nc)+{\rm Tr}_k\left(\frac{c}{Z_n(v)}\;{\rm Tr}_k^{nk}(B_n(v)+Z_n(v))\right)\\
&=&{\rm Tr}_k(nc)\enspace.
\end{eqnarray*}
 \setlength{\arraycolsep}{5pt}\noindent
This completes the proof.
\hspace*{\fill}${\Box}$

\begin{proposition}
 \label{pr:A2kZero}
Let $n$ be odd and take any $a\in\mathrm{GF}(2^{nk})^*$. Then
polynomial $A_a(x)$ has exactly $2^k$ zeros in $\mathrm{GF}(2^{nk})$
if and only if $Z_n(a)=0$ and $B_n(a)\neq 0$. Moreover, these zeros
are the following
\[v_\mu=c\sum_{i=0}^{\frac{n-1}{2}}\frac{B_{n-1}^{2^{(2i+1)k}}(a)}{B_n^{2^{(2i+1)k}+2^{2ik}-1}(a)}+\mu B_n(a)\]
with $\mu\in\mathrm{GF}(2^k)$ and for each zero of this type ${\rm
Tr}_{nk}(v_\mu)=0$. Also $|M_{2^k}|=2^{(n-1)k}-1$.
\end{proposition}

{\bf Proof.} First, we consider $l_a(x)=a_1 x^{2^{2k}}+x^{2^k}+a_0
x$ being the linearized homogeneous part of $A_a(x)$, and prove that
it has exactly $2^k$ zeros in $\mathrm{GF}(2^{nk})$ if and only if
$Z_n(a)=0$ and $B_n(a)\neq 0$.

Assume that $Z_n(a)=0$ and $B_n(a)\neq 0$. Then, by (\ref{eq:A(B)}),
\begin{equation}
 \label{eq:Bhom}
a_1 B_n^{2^{2k}}(a)+B_n^{2^k}(a)+a_0 B_n(a)=0
\end{equation}
which means that all $2^k$ distinct values $\mu B_n(a)$ for
$\mu\in\mathrm{GF}(2^k)$ are zeros of $l_a(x)$. It is not difficult
to see that in our case $l_a(x)$ can not have more than $2^k$ zeros
in $\mathrm{GF}(2^{nk})$. Indeed, consider matrix $\mathcal{M}_n$ of
the system of $n$ linear equations (\ref{eq:syst}) with $c=0$. Note
that $\det\mathcal{M}_n=Z_n^2(a)=0$ and a principal submatrix
obtained by deleting the last column and the last row from
$\mathcal{M}_n$ is nonsingular with the determinant
$\Delta_a(1,n-2)=B_n^2(a)\neq 0$ (see (\ref{eq:Delta})). Therefore,
applying equivalent row transformations to $\mathcal{M}_n$ we can
obtain a matrix containing a nonsingular diagonal submatrix lying in
the first $n-1$ columns and rows. Thus, the equation given by the
first row of this equivalent matrix is nonzero and has degree $2^k$.
We conclude that homogeneous system (\ref{eq:syst}) can not have
more than $2^k$ solutions and the same holds for the equation
$l_a(x)=0$.

Now we prove the converse implication. Assume that $l_a(x)$ has
exactly $2^k$ zeros in $\mathrm{GF}(2^{nk})$. Here we use the
technique found by Bluher \cite{Bl04} for counting the number of
$b\in\mathrm{GF}(2^{nk})^*$ for which $f_b(y)=y^{2^k+1}+by+b$ has
exactly one zero in $\mathrm{GF}(2^{nk})$. Denote
$S=\{x\in\mathrm{GF}(2^{nk})\setminus\mathrm{GF}(2^k)\ |\ {\rm
Tr}^{nk}_k(x)\neq 0\}$. For any $v\in S$ define
$r=v^{1-2^k}+1\in\mathrm{GF}(2^{nk})\setminus\{0,1\}$ and
corresponding $b=\frac{r^{2^k+1}}{r+1}\neq 0$. Obviously, such an
$r$ is a zero of $f_b(y)$. Note that
\begin{equation}
 \label{eq:b1}
b=\frac{r^{2^k+1}}{r+1}=v^{2^k-1}(v^{1-2^k}+1)^{2^k+1}=
\frac{(v+v^{2^k})^{2^k+1}}{v^{2^{2k}+1}}=V^{-1}\enspace,
\end{equation}
where $V$ comes from (\ref{eq:V}). Then, by Lemma~\ref{le:ZeB} and
Corollary~\ref{co:ZeZ}, $B_n(b^{-1})\neq 0$ and $Z_n(b^{-1})=0$. By
the implication already proved, $l_{b^{-1}}(x)$ has exactly $2^k$
zeros in $\mathrm{GF}(2^{nk})$. Multiplying the latter polynomial by
$b_0 b_1 x^{-1}$ (we remove one zero $x=0$) and then substituting
$x$ with $b_0 y$ leads to $b_1^{2^k} y^{2^{2k}-1}+b_1^2
y^{2^k-1}+b_1$ with $2^k-1$ zeros and having the form of polynomial
$g(x)$ from Theorem~\ref{th:Bl}. Thus, by (\ref{it:2}) in this
theorem, $f_b(y)$ (as well as $f_{b^{2^{l+1}}}(y)$) has exactly one
zero in $\mathrm{GF}(2^{nk})$.

Now we prove that function (\ref{eq:b1}) that maps every $v\in S$ to
$b\in\mathrm{GF}(2^{nk})^*$ is a $(2^k-1)$-to-$1$ mapping. First,
note that $(2^k-1)$-power is a $(2^k-1)$-to-$1$ mapping of $S$ to
$\mathrm{GF}(2^{nk})^*$. Indeed, if $x\in S$ and $x^{2^k-1}=t$ then
the latter identity holds for all distinct $\delta x\in S$ with
$\delta\in\mathrm{GF}(2^k)^*$ since ${\rm Tr}^{nk}_k(\delta
x)=\delta{\rm Tr}^{nk}_k(x)\neq 0$ and $\delta
x\notin\mathrm{GF}(2^k)$. Thus, every $r=v^{1-2^k}+1$ is obtained
from $2^k-1$ different values of $v$. Finally, the mapping from $r$
to $b$ is $1$-to-$1$ since for the obtained $b$ the equation
$f_b(y)=0$ has exactly one root $r$.

Therefore, taking all $v\in S$ and using (\ref{eq:b1}), we obtain
$|S|/(2^k-1)$ different values of $b\in\mathrm{GF}(2^{nk})^*$ and
this number is equal to the total number of $b$ such that $f_b(y)$
has exactly one zero (see Theorem~\ref{th:Bl}). Therefore, these and
only these values of $b$ satisfying (\ref{eq:b1}) result in the
polynomials $f_b(y)$ having exactly one zero.

Dividing $l_a(x)$ by $a_0 a_1 x$ (we remove one zero $x=0$) and then
substituting $x$ with $a_0^{-1}y$ leads to $a_1^{-2^k}
y^{2^{2k}-1}+a_1^{-2} y^{2^k-1}+a_1^{-1}$ which has the form of
polynomial $g(x)$ from Theorem~\ref{th:Bl} taking $b=a_1^{-1}$.
Thus, $f_{a^{-1}}(y)$ (as well as $f_{a^{-2^{l+1}}}(y)$) has exactly
one zero in $\mathrm{GF}(2^{nk})$ and $a^{-1}$ is obtained by
(\ref{eq:b1}). Therefore, by Lemma~\ref{le:ZeB} and
Corollary~\ref{co:ZeZ}, $B_n(a)\neq 0$ and $Z_n(a)=0$.

If $A_a(x)$ has exactly $2^k$ zeros in $\mathrm{GF}(2^{nk})$ then
the same holds for its homogeneous part $l_a(x)$ and we already
proved that in this case, $Z_n(a)=0$ and $B_n(a)\neq 0$. Now we have
to find a particular solution of $A_a(x)=0$ assuming $Z_n(a)=0$ and
$B_n(a)\neq 0$. By Corollary~\ref{co:ZeZ}, $a$ has the form of
(\ref{eq:V}) and, by Lemma~\ref{le:ZeB}, ${\rm Tr}^{nk}_k(v_0)\neq
0$. Using these facts and assuming that $n$ is odd (note that the
latter assumption is involved only at this stage), we compute
 \setlength{\arraycolsep}{0.0em}
\begin{eqnarray*}
&&A_a\left(c\sum_{i=0}^{\frac{n-1}{2}}\frac{B_{n-1}^{2^{(2i+1)k}}(a)}{B_n^{2^{(2i+1)k}+2^{2ik}-1}(a)}\right)=
c a_1 B_n^{2^{2k}}(a)\sum_{i=1}^{\frac{n-1}{2}}\frac{B_{n-1}^{2^{(2i+1)k}}(a)}{B_n^{2^{(2i+1)k}+2^{2ik}}(a)}\\
&&\quad+\:c a_1 B_n^{2^{2k}}(a)\frac{B_{n-1}^{2^{2k}}(a)}{B_n^{2^{2k}+2^k}(a)}+
c B_n^{2^k}(a)\sum_{i=1}^{\frac{n-1}{2}}\frac{B_{n-1}^{2^{2ik}}(a)}{B_n^{2^{2ik}+2^{(2i-1)k}}(a)}\\
&&\quad+\:c B_n^{2^k}(a)\frac{B_{n-1}^{2^k}(a)}{B_n^{2^k+1}(a)}+
c a_0 B_n(a)\sum_{i=0}^{\frac{n-1}{2}}\frac{B_{n-1}^{2^{(2i+1)k}}(a)}{B_n^{2^{(2i+1)k}+2^{2ik}}(a)}+c\\
&&\stackrel{(\ref{eq:Bhom})}{=}
\:c B_n^{2^k}(a)\sum_{i=0}^{\frac{n-1}{2}}\frac{B_{n-1}^{2^{(2i+1)k}}(a)}{B_n^{2^{(2i+1)k}+2^{2ik}}(a)}+
c B_n^{2^k}(a)\sum_{i=1}^{\frac{n-1}{2}}\frac{B_{n-1}^{2^{2ik}}(a)}{B_n^{2^{2ik}+2^{(2i-1)k}}(a)}\\
&&\quad+\:c a_0 B_n(a)\frac{B_{n-1}^{2^{k}}(a)}{B_n^{2^{k}+1}(a)}+
c a_1 B_n^{2^{2k}}(a)\frac{B_{n-1}^{2^{2k}}(a)}{B_n^{2^{2k}+2^k}(a)}+c\\
&&=\:c B_n^{2^k}(a){\rm Tr}_k^{nk}\left(\frac{B_{n-1}^{2^k}(a)}{B_n^{2^k+1}(a)}\right)+
c\frac{a_0 B_{n-1}^{2^{k}}(a)+(a_0 B_{n-1}^{2^k}(a))^{2^k}}{B_n^{2^k}(a)}+c\\
&&\stackrel{(\ast)}{=}\:c\frac{(a_{n-1}
B_{n-1}(a))^{2^{k}}+B_{n+1}^{2^k}(a)}{B_n^{2^k}(a)}+c\stackrel{(\ref{eq:dB1})}{=}\;0\enspace,
\end{eqnarray*}
 \setlength{\arraycolsep}{5pt}\noindent
where $(\ast)$ holds by Corollary~\ref{co:ZeTr} and since
$B_{n+1}(a)=a_0 B_{n-1}^{2^k}(a)$ resulting from (\ref{eq:Z}) if
$Z_n(a)=0$.

Finally, to prove the trace identity for $v_\mu$ first note that, by
(\ref{eq:TrBZ}), ${\rm Tr}_k^{nk}(B_n(a)+Z_n(a))={\rm
Tr}_k^{nk}(B_n(a))=0$ if $Z_n(a)=0$. Further,
 \setlength{\arraycolsep}{0.0em}
\begin{eqnarray*}
&&{\rm Tr}_k^{nk}\left(\sum_{i=0}^{\frac{n-1}{2}}\frac{B_{n-1}^{2^{(2i+1)k}}(a)}{B_n^{2^{(2i+1)k}+2^{2ik}-1}(a)}\right)
=\sum_{j=0}^{n-1}\sum_{i=0}^{\frac{n-1}{2}}
\frac{B_n^{2^{jk}}(a)B_{n-1}^{2^{(2i+j+1)k}}(a)}{B_n^{2^{(2i+j+1)k}+2^{(2i+j)k}}(a)}\\
&&=\:\sum_{j=0}^{n-1}\frac{B_{n-1}^{2^{(j+1)k}}(a)\sum_{i=0}^{\frac{n-1}{2}}B_n^{2^{(j-2i)k}}(a)}{B_n^{2^{(j+1)k}+2^{jk}}(a)}
={\rm Tr}_k^{nk}\left(\frac{B_{n-1}^{2^k}(a)\sum_{i=0}^{\frac{n-1}{2}}B_n^{2^{(2i+1)k}}(a)}{B_n^{2^{k}+1}(a)}\right)\\
&&=\:{\rm Tr}_k^{nk}\left(\frac{B_{n-1}^{2^k}(a)\sum_{i=0}^{\frac{n-1}{2}}\left(B_{n+1}(a)+
B_{n+1}^{2^k}(a)\right)^{2^{2ik}}}{B_n^{2^{k}+1}(a)}\right)\\
&&=\:{\rm Tr}_k^{nk}\left(\frac{B_{n-1}^{2^k}(a)\left({\rm Tr}_k^{nk}(B_{n+1}(a))+B_{n+1}(a)\right)}{B_n^{2^{k}+1}(a)}\right)\\
&&=\:{\rm Tr}_k^{nk}(B_{n+1}(a)){\rm Tr}_k^{nk}\left(\frac{B_{n-1}^{2^k}(a)}{B_n^{2^{k}+1}(a)}\right)+
{\rm Tr}_k^{nk}\left(\frac{B_{n-1}^{2^k}(a)B_{n+1}(a)}{B_n^{2^{k}+1}(a)}\right)=0\enspace,
\end{eqnarray*}
 \setlength{\arraycolsep}{5pt}\noindent
where the latest identity follows by Corollary~\ref{co:ZeTr}.

The identity for $|M_{2^k}|$ follows from Theorem~\ref{th:Bl}.
\hspace*{\fill}${\Box}$

Now we are left with the remaining case when $B_n(a)=0$ (then,
obviously, $Z_n(a)=0$). In the following proposition, the ``only if"
part follows from Propositions~\ref{pr:A1Zero} and \ref{pr:A2kZero}.
We provide this proof yet, independently of previous statements,
since its major part contains the result used for proving the
converse implication and also needed for proving the fact from
Lemma~\ref{le:ZeB}.

\begin{proposition}
 \label{pr:l22kZero}
Take any $a\in\mathrm{GF}(2^{nk})^*$. Then polynomial $l_a(x)=a_1
x^{2^{2k}}+x^{2^k}+a_0 x$ has exactly $2^{2k}$ zeros in
$\mathrm{GF}(2^{nk})$ if and only if $B_n(a)=0$.
\end{proposition}

{\bf Proof.} Assume that $l_a(x)$ has exactly $2^{2k}$ zeros in
$\mathrm{GF}(2^{nk})$. Here we use the technique found by Bluher
\cite{Bl04} for counting the number of $b\in\mathrm{GF}(2^{nk})^*$
for which $f_b(y)=y^{2^k+1}+by+b$ has $2^k+1$ zeros in
$\mathrm{GF}(2^{nk})$. Denote
$G=\mathrm{GF}(2^{nk})\setminus\mathrm{GF}(2^{2k})$. Take any
$u\in\mathrm{GF}(2^{nk})$ such that $u\notin\mathrm{GF}(2^{2k})$
which implies $u^{2^{2k}}\neq u$ and $(u+u^{2^k})^{2^k}\neq
u+u^{2^k}$ or, equivalently, $u+u^{2^k}\notin\mathrm{GF}(2^k)$. Now
we can define
$r=(u+u^{2^k})^{1-2^k}+1\in\mathrm{GF}(2^{nk})\setminus\{0,1\}$ and
corresponding $b=\frac{r^{2^k+1}}{r+1}\neq 0$. Obviously, such an
$r$ is a zero of $f_b(y)$. Define also $r_0=ru^{2^k-1}$ and
$r_1=r(u+1)^{2^k-1}$ and note that $r$, $r_0$ and $r_1$ are pairwise
distinct. Further,
\[f_b(r_0)=r^{2^k+1}u^{2^{2k}-1}+bru^{2^k-1}+b=b((r+1)u^{2^{2k}}+ru^{2^k}+u)/u=0\]
since $r(u+u^{2^k})^{2^k}=u+u^{2^{2k}}$ by the definition of $r$.
Also, similarly, we get
\[f_b(r_1)=b((r+1)(u+1)^{2^{2k}}+r(u+1)^{2^k}+(u+1))/(u+1)=b(r+1+r+1)/(u+1)=0\ \ .\]
Thus, $f_b(y)$ with such a $b$ has at least three zeros and, by
Theorem~\ref{th:Bl}, it has $2^k+1$ zeros. Note that
\begin{equation}
 \label{eq:b}
b=\frac{r^{2^k+1}}{r+1}=(u+u^{2^k})^{2^k-1}((u+u^{2^k})^{1-2^k}+1)^{2^k+1}=
\frac{(u+u^{2^{2k}})^{2^k+1}}{(u+u^{2^k})^{2^{2k}+1}}=V^{-1}\enspace,
\end{equation}
where $V$ comes from (\ref{eq:V}) assuming $v=u+u^{2^k}$.

Now we prove that function (\ref{eq:b}) that maps every $u\in G$ to
$b\in\mathrm{GF}(2^{nk})^*$ is a $(2^{3k}-2^k)$-to-$1$ mapping.
First, note that $u+u^{2^k}$ is a $2^k$-to-$1$ mapping onto
$F=\{x\in\mathrm{GF}(2^{nk})\setminus\mathrm{GF}(2^k)\ |\ {\rm
Tr}^{nk}_k(x)=0\}$. Further, $(2^k-1)$-power is a $(2^k-1)$-to-$1$
mapping of $F$ to $\mathrm{GF}(2^{nk})^*$. Indeed, if $x\in F$ and
$x^{2^k-1}=t$ then the latter identity holds for all distinct
$\delta x\in F$ with $\delta\in\mathrm{GF}(2^k)^*$ since ${\rm
Tr}^{nk}_k(\delta x)=\delta{\rm Tr}^{nk}_k(x)=0$ and $\delta
x\notin\mathrm{GF}(2^k)$. Thus, every $r=(u+u^{2^k})^{1-2^k}+1$ is
obtained from $2^k(2^k-1)$ different values of $u$. Finally, the
mapping from $r$ to $b$ is $(2^k+1)$-to-$1$ since for the obtained
$b$ the equation $f_b(y)=0$ has $(2^k+1)$ roots and every root $r$
satisfies $(r+1)^{-1}\in F^{2^k-1}$. Indeed, let $r$, $r_0$ and
$r_1$ be any distinct zeros of $f_b(y)$ (not necessarily the ones
defined above) and define $u=(r+r_1)/(r_0+r_1)$. Note that
\[rr_0(r+r_0)^{2^k}=r_0 r^{2^k+1}+r r_0^{2^k+1}=r_0 b(r+1)+r b(r_0+1)=b(r+r_0)\]
and so $b=rr_0(r+r_0)^{2^k-1}=rr_1(r+r_1)^{2^k-1}=r_0
r_1(r_0+r_1)^{2^k-1}$. Then
\[u^{2^{2k}-1}=(r_0/r)^{2^k+1}=(r_0+1)/(r+1)\neq 1\]
and thus, $u\in G$. The identity $(r+1)^{-1}=(u+u^{2^k})^{2^k-1}\in
F^{2^k-1}$ follows from \cite[Lemma~2.1]{Bl04}.

Therefore, taking all $u\in G$ and using (\ref{eq:b}), we obtain
$|G|/(2^{3k}-2^k)$ different values of $b\in\mathrm{GF}(2^{nk})^*$
and this number is equal to the total number of $b$ such that
$f_b(y)$ has $2^k+1$ zeros (see Theorem~\ref{th:Bl}). Therefore,
these and only these values of $b$ satisfying (\ref{eq:b}) result in
the polynomials $f_b(y)$ having $2^k+1$ zeros.

Dividing $l_a(x)$ by $a_0 a_1 x$ (we remove one zero $x=0$) and then
substituting $x$ with $a_0^{-1}y$ leads to $a_1^{-2^k}
y^{2^{2k}-1}+a_1^{-2} y^{2^k-1}+a_1^{-1}$ which has the form of
polynomial $g(x)$ from Theorem~\ref{th:Bl} taking $b=a_1^{-1}$.
Thus, $f_{a^{-1}}(y)$ (as well as $f_{a^{-2^{k+1}}}(y)$) has exactly
$2^k+1$ zeros in $\mathrm{GF}(2^{nk})$ and $a^{-1}$ is obtained by
(\ref{eq:b}). Therefore, by Lemma~\ref{le:ZeB}, $B_n(a)=0$ .

The converse implication is easy now. If $B_n(a)=0$ then, by
Lemma~\ref{le:ZeB}, $a$ has the form of (\ref{eq:V}) with
$v=u+u^{2^k}$ for some $u\in G$. Then the corresponding $b=a^{-1}$
has the form of (\ref{eq:b}) and, by the fact proved above, the
polynomial $f_b(y)$ has $2^k+1$ zeros which, by
Theorem~\ref{th:Bl}~(\ref{it:3}), is equivalent to $l_a(x)$ having
$2^{2k}$ zeros. \hspace*{\fill}${\Box}$

In the following proposition, we prove that $A_a(x)=0$ always has a
solution if $n$ is odd and ${\rm Tr}_k(c)=1$ (which is also valid
for even $n$ but this case is not relevant to the current paper).

\begin{proposition}
 \label{pr:A22kZero}
Take any $a\in\mathrm{GF}(2^{nk})$, where $n$ is odd and
$c\in\mathrm{GF}(2^k)$ with ${\rm Tr}_k(c)=1$. Then polynomial
$A_a(x)$ has at least one zero in $\mathrm{GF}(2^{nk})$. Moreover,
if $A_a(x)$ has exactly $2^{2k}$ zeros then ${\rm Tr}_{nk}(v)=1$ for
any $v\in\mathrm{GF}(2^{nk})$ with $A_a(v)=0$ and
$|M_{2^{2k}}|=\frac{2^{(n-1)k}-1}{2^{2k}-1}$.
\end{proposition}

{\bf Proof.} Take any pair
$(a,v)\in\mathrm{GF}(2^{nk})\times\mathrm{GF}(2^{nk})$ with $v\neq
c$ such that $A_a(v)=0$. Then, assuming
$b=a\left(\frac{v}{v+c}\right)^{2^k+1}$, we obtain
 \setlength{\arraycolsep}{0.14em}
\begin{eqnarray*}
A_b(v+c)&=&a^{2^k}\frac{v^{2^{2k}+2^k}}{(v+c)^{2^k}}+(v+c)^{2^k}+a\frac{v^{2^k+1}}{(v+c)^{2^k}}+c\\
&=&\frac{1}{(v+c)^{2^k}}\left(v^{2^k}(a^{2^k}
v^{2^{2k}}+v^{2^k}+av)+c^2+c(v^{2^k}+c)\right)=0\enspace.
\end{eqnarray*}
 \setlength{\arraycolsep}{5pt}\noindent
Since $n$ is odd and ${\rm Tr}_k(c)=1$, we obtain a $1$-to-$1$
correspondence between two sets
 \setlength{\arraycolsep}{0.14em}
\begin{eqnarray*}
S_0&=&\{(a,v)\;|\;v\neq c, A_a(v)=0, {\rm Tr}_{nk}(v)=0\}\quad\mbox{and}\\
S_1&=&\{(a,v)\;|\;v\neq c, A_a(v)=0, {\rm Tr}_{nk}(v)=1\}
\end{eqnarray*}
 \setlength{\arraycolsep}{5pt}\noindent
defined by $(a,v)\mapsto(b,v+c)$ with
$b=a\left(\frac{v}{v+c}\right)^{2^k+1}$ and thus, $|S_0|=|S_1|$.
Note that $A_a(c)=c(a^{2^k}+a)=0$ if and only if
$a\in\mathrm{GF}(2^k)$. Now, since $A_a(x)$ can have $0$, $1$, $2^k$
or $2^{2k}$ zeros, we can compute the following sum in two different
ways
 \setlength{\arraycolsep}{0.0em}
\begin{eqnarray*}
&&\sum_{(a,v):\,A_a(v)=0}(-1)^{{\rm Tr}_{nk}(v)}=|S_0|-|S_1|-2^k\\
&&=\:(-1)^{{\rm Tr}_{nk}(c)}+\sum_{a\in M_1}(-1)^{{\rm
Tr}_{nk}(\mathcal{V}_a)}+
\sum_{a\in M_{2^k}}\sum_{v:\,A_a(v)=0}(-1)^{{\rm Tr}_{nk}(v)}+X\\
&&=\:-1-|M_1|+2^k|M_{2^k}|+X\enspace,
\end{eqnarray*}
 \setlength{\arraycolsep}{5pt}\noindent
by Propositions~\ref{pr:A1Zero} and \ref{pr:A2kZero}, where
$X=\sum_{a\in M_{2^{2k}}}\sum_{v:\,A_a(v)=0}(-1)^{{\rm
Tr}_{nk}(v)}$. Then
\[X=-\frac{2^{2k}(2^{(n-1)k}-1)}{2^{2k}-1}=-2^{2k}(|\mathrm{GF}(2^{nk})^*|-|M_1|-|M_{2^k}|)\]
which holds if and only only if
$|M_{2^{2k}}|=\frac{2^{(n-1)k}-1}{2^{2k}-1}$ and ${\rm
Tr}_{nk}(v)=1$ for any $v\in\mathrm{GF}(2^{nk})$ with $A_a(v)=0$ and
$a\in M_{2^{2k}}$. Since
$|M_1|+|M_{2^k}|+|M_{2^{2k}}|=|\mathrm{GF}(2^{nk})^*|$ and $A_0(x)$
has a unique zero $x=c$, polynomial $A_a(x)$ has at least one zero
in $\mathrm{GF}(2^{nk})$ for any $a\in\mathrm{GF}(2^{nk})$.
\hspace*{\fill}${\Box}$

\section{The Linearized Polynomial $L_a(z)$}
 \label{sec:Lin}
The distribution of the four-valued cross-correlation function to be
determined in Section~\ref{sec:4VDist} depends on the detailed
distribution of the number of zeros in $\mathrm{GF}(2^{2nk})$, with
$n>2$, of the linearized polynomial
\begin{equation}
 \label{eq:L}
L_a(z)=z^{2^{(n+1)k}}+r^{2^k}a^{2^k}z^{2^{2k}}+raz\enspace,
\end{equation}
where $n$ is odd, $a\in\mathrm{GF}(2^{nk})$ and
$r\in\mathrm{GF}(2^{2nk})^*$ with $r^{2^{nk}+1}=1$ but
$r^{\frac{2^{nk}+1}{2^k+1}}\neq 1$. For the details on linearized
polynomials in general, the reader is referred to Lidl and
Niederreiter \cite{LiNi97}. It is clear that $L_a(z)$ does not have
multiple roots if $a\neq 0$. In the following propositions, we
always take $L_a(z)$ defined in (\ref{eq:L}).

Define polynomials $Y_n(x)$ over $\mathrm{GF}(2^{nk})$ as
\[Y_n(x)=Z^2_n(x)+{\rm N}^{nk}_k(x)(\delta+\delta^{-1})\enspace,\]
where $\delta=r^{\frac{2^{nk}+1}{2^k+1}}\in\mathrm{GF}(2^{2k})$ is a
$(2^k+1)^{\rm{th}}$ root of unity over $\mathrm{GF}(2)$ and $Z_n(x)$
comes from (\ref{eq:Z}). Also, for any $v\in\mathrm{GF}(2^{2nk})$
denote $v_i=v^{2^{ik}}$ for $i\geq 0$ so $L_a(z)=z_{n+1}+r_1 a_1
z_2+r_0 a_0 z_0$. Finally, for $0<j\leq i$ and
$x\in\mathrm{GF}(2^{nk})$, let $D^{j,i}_x$ denote a three-diagonal
matrix of size $i-j+2$ that contains ones on the main diagonal and
with
\[D^{j,i}_x(t,t+1)=r^{(-1)^{j+t-1}}_{j+t}x_{j+t}\quad\mbox{and}\quad
D^{j,i}_x(t+1,t)=r^{(-1)^{j+t}}_{j+t}x_{j+t}\] for $t=0,\dots,i-j$,
where the indices of $r$, $x$ and powers of $r$ are reduced modulo
$n$ (the only exception is $j+t=n$ when
$r^{(-1)^{n-1}}_n=r^{-1}_0$), rows and columns of $D^{j,i}_x$ are
numbered from $0$ to $i-j+1$. The determinant of $D^{j,i}_x$,
denoted as $\Delta'_x(j,i)$, can be computed expanding by minors
along the last row to obtain
\[\Delta'_x(j,i)=\Delta'_x(j,i-1)+x_i^2\Delta'_x(j,i-2)\]
assuming $\Delta'_x(j,i)=1$ if $i-j\in\{-2,-1\}$. Comparing the
latter recursive identity with (\ref{eq:RecDelta}) it is easy to see
that
\begin{equation}
 \label{eq:DeltaPr}
\Delta'_x(j,i)=\Delta_x(j,i)\enspace.
\end{equation}

Zeros of $L_a(z)$ in $\mathrm{GF}(2^{2nk})$ form a vector space over
$\mathrm{GF}(2^{2k})$. Since the degree of $L_a(z)$ is $2^{(n+1)k}$,
the number of zeros is at most $2^{(n+1)k}$, and thus, the dimension
of the vector space over $\mathrm{GF}(2^{2k})$ is at most $(n+1)/2$.
Therefore, $L_a(z)$ has either $1,2^{2k},2^{4k},\dots,2^{(n+1)k}$
zeros in $\mathrm{GF}(2^{2nk})$. However, in
Proposition~\ref{pr:Lle22kZero} we prove that $L_a(z)$ can not have
more than $2^{2k}$ zeros in $\mathrm{GF}(2^{2nk})$ and thus, the
only possibilities are either $1$ or $2^{2k}$ zeros.

\begin{proposition}
 \label{pr:L1Zero}
For any $a\in\mathrm{GF}(2^{nk})^*$, if $Y_n(a)\neq 0$ then
$L_a(z)=0$ has exactly one root in $\mathrm{GF}(2^{2nk})$ that is
equal to zero. In particular, if $Z_n(a)=0$ ($Z_n(x)$ defined in
(\ref{eq:Z})) then $L_a(z)$ has exactly one zero.
\end{proposition}

{\bf Proof.} Obviously, $L_a(0)=0$ and we have to show that this is
the only zero of $L_a(z)$ in $\mathrm{GF}(2^{2nk})$ if $Y_n(a)\neq
0$. Taking equation $L_a(z)=0$ and all its $2^{2ik}$ powers we
obtain $n$ equations
\[L_a^{2^{2ik}}(x)=z_{n+2i+1}+r_{2i+1}a_{2i+1}z_{2i+2}+r_{2i}a_{2i}z_{2i}=0\quad\mbox{for}\quad
i=0,\dots,n-1\enspace,\] where all indices are calculated modulo
$2n$. If $z_{2i}$ $(i=0,\dots,n-1)$ are considered as independent
variables then matrix $\mathcal{M}_n$ of the obtained system of $n$
linear equations with $n$ unknowns consists of three cyclic
antidiagonals and
 \setlength{\arraycolsep}{0.14em}
\begin{eqnarray*}
\mathcal{M}_n(i,(n-3)/2-i)&=&1\ ,\\
\mathcal{M}_n(i,n-i-1)&=&r_{2i}a_{2i}\ ,\\
\mathcal{M}_n(i,n-i-2)&=&r_{2i+1}a_{2i+1}\quad\mbox{for}\quad
i=0,\dots,n-1\enspace,
\end{eqnarray*}
 \setlength{\arraycolsep}{5pt}\noindent
where rows and columns of $\mathcal{M}_n(i,j)$ are numbered from $0$
to $n-1$ and all elements of $\mathcal{M}_n$ are indexed modulo $n$.

Now permute the columns and rows of $\mathcal{M}_n$ in the following
way. Decimate the rows as $i(n+1)/2$ and columns as
$(n-3)/2+i(n-1)/2$ modulo $n$ for $i=0,\dots,n-1$ (note that
$\gcd((n+1)/2,n)=\gcd((n-1)/2,n)=1$). Then the obtained matrix
$\mathcal{M}'_n$ is three-diagonal cyclic with
 \setlength{\arraycolsep}{0.14em}
\begin{eqnarray*}
\mathcal{M}'_n(i,i)&=&\mathcal{M}_n(i(n+1)/2,(n-3)/2+i(n-1)/2)=1\ ,\\
\mathcal{M}'_n(i,i-1)&=&\mathcal{M}_n(i(n+1)/2,(n-3)/2+(i-1)(n-1)/2)=r_{i(n+1)}a_{i(n+1)}\ ,\\
\mathcal{M}'_n(i,i+1)&=&\mathcal{M}_n(i(n+1)/2,(n-3)/2+(i+1)(n-1)/2)=r_{i(n+1)+1}a_{i(n+1)+1}
\end{eqnarray*}
 \setlength{\arraycolsep}{5pt}\noindent
for $i=0,\dots,n-1$ (indices of $r$ and $a$ are calculated modulo
$2n$) since
 \setlength{\arraycolsep}{0.0em}
\begin{eqnarray*}
&&i(n+1)/2+(n-3)/2+i(n-1)/2=(n-3)/2+in\equiv (n-3)/2\ (\bmod\;n)\ ,\\
&&i(n+1)/2+(n-3)/2+(i-1)(n-1)/2=-1+in\equiv n-1\ (\bmod\;n)\ ,\\
&&i(n+1)/2+(n-3)/2+(i+1)(n-1)/2=n-2+in\equiv n-2\
(\bmod\;n)\enspace.
\end{eqnarray*}
 \setlength{\arraycolsep}{5pt}\noindent
Also note that $a_{i(n+1)}=a_i$ since $a\in\mathrm{GF}(2^{nk})$ and
$r_{i(n+1)}=r^{(-1)^i}_i$ since $r_{n+i}=r^{2^{ik}}_n=r^{-1}_i$ for
any $i\geq 0$. Then for $i=0,\dots,n-1$
\[\mathcal{M}'_n(i,i+1)=r^{(-1)^i}_{i+1} a_{i+1}\quad\mbox{and}
\quad\mathcal{M}'_n(i+1,i)=r^{(-1)^{i+1}}_{i+1} a_{i+1}\quad\mbox{so}\]
\begin{equation*}
\mathcal{M}'_n=\left(\begin{array}{cccccc}
1&r_1 a_1&0&\cdots&r_0 a_0\\
r^{-1}_1 a_1&\ddots&\ddots&\ddots&0\\
\vdots&\ddots&\ddots&\ddots&\vdots\\
0&&\ddots&1&r^{-1}_{n-1} a_{n-1}\\
r^{-1}_0 a_0&0&\cdots&r_{n-1} a_{n-1}&1
\end{array}\right)\enspace.
\end{equation*}
Note that a principal submatrix obtained by deleting the last column
and the last row from $\mathcal{M}'_n$ is exactly $D^{1,n-2}_a$.

We also have to apply the decimation $(n-3)/2+i(n-1)/2$ modulo $n$
for $i=0,\dots,n-1$ (used to permute the columns of $\mathcal{M}$)
to the vector of unknowns $(z_{2(n-1)},z_{2(n-2)},\dots,z_2,z_0)$.
This results in
$\vec{z}=(z_{n+1},z_2,z_{n+3},\dots,z_{n-1},z_0)^{\rm T}$, where the
increment for the index of $z$ is equal to $n-1$ starting from $0$
and going right to left (indices are calculated modulo $2n$). Now,
if $\vec{0}=(0,\dots,0)^{\rm T}$ then a new system has the following
matrix representation
\begin{equation}
 \label{eq:syst1}
\mathcal{M}'_n\vec{z}=\vec{0}\enspace.
\end{equation}
The determinant of $\mathcal{M}_n$ is equal to the determinant of
$\mathcal{M}'_n$ and can be computed expanding the latter by minors
along the last row. Doing this it is easy to see that
 \setlength{\arraycolsep}{0.14em}
\begin{eqnarray*}
\det\mathcal{M}'_n&=&\Delta'_a(1,n-2)+r_{n-1}a_{n-1}
\left(r^{-1}_{n-1}a_{n-1}\Delta'_a(1,n-3)+\prod_{i=0}^{n-2}r^{(-1)^i}_i a_i\right)\\
&&\quad\quad\quad\quad\quad\ \ +\;r^{-1}_0 a_0\left(r_0 a_0\Delta'_a(2,n-2)+\prod_{i=1}^{n-1}r^{(-1)^{i-1}}_i a_i\right)\\
&\stackrel{(\ref{eq:Delta},\ref{eq:Delta2k},\ref{eq:DeltaPr})}{=}&B_n^2(a)+a_{n-1}^2
B_{n-1}^2(a)+(a_0 B_{n-1}^{2^k}(a))^2+{\rm
N}^{nk}_k(a)(\delta+\delta^{-1})\\
&\stackrel{(\ref{eq:dB1},\ref{eq:Z})}{=}&Z^2_n(a)+{\rm
N}^{nk}_k(a)(\delta+\delta^{-1})=Y_n(a)\enspace.
\end{eqnarray*}
 \setlength{\arraycolsep}{5pt}\noindent
Thus, if $Y_n(a)\neq 0$ then (\ref{eq:syst1}) has only zero
solution. Now note that every $v\in\mathrm{GF}(2^{2nk})$ with
$L_a(v)=0$ provides a solution to the system given by
$v_{2i}=v^{2^{2ik}}$ for $i=0,\dots,n-1$. Therefore, if $Y_n(a)\neq
0$ then $L_a(z)$ has at most one zero.

Finally, note that if $Z_n(a)=0$ (obviously, $a\neq 0$) then
$Y_n(a)={\rm N}^{nk}_k(a)(\delta+\delta^{-1})\neq 0$ and thus,
$L_a(z)$ has exactly one zero. \hspace*{\fill}${\Box}$

\begin{proposition}
 \label{pr:Lle22kZero}
For any $a\in\mathrm{GF}(2^{nk})^*$, $L_a(z)$ has at most $2^{2k}$
zeros in $\mathrm{GF}(2^{2nk})$.
\end{proposition}

{\bf Proof.} Consider the homogeneous system of linear equations
(\ref{eq:syst1}) defined by matrix $\mathcal{M}'_n$ with
$\vec{z}=(z_{n+1},z_2,z_{n+3},\dots,z_{n-1},z_0)^{\rm T}$. Note that
a principal submatrix obtained by deleting the last column and the
last row from $\mathcal{M}'_n$ is exactly $D^{1,n-2}_a$ and
\[\det D^{1,n-2}_a=\Delta'_a(1,n-2)\stackrel{(\ref{eq:DeltaPr})}{=}\Delta_a(1,n-2)
\stackrel{(\ref{eq:Delta})}{=}B^2_n(a)\enspace.\] After removing the
last equation from (\ref{eq:syst1}), we can write the remaining
system as
\begin{equation}
 \label{eq:SystRed}
D^{1,n-2}_a\vec{z}'=(r_0 a_0
z_0,0,\dots,0,r^{-1}_{n-1}a_{n-1}z_0)^{\rm T}\enspace,
\end{equation}
where $\vec{z}'=(z_{n+1},z_2,z_{n+3},\dots,z_{n-1})^{\rm T}$ is
obtained from $\vec{z}$ by deleting the last coordinate $z_0$.

Let $\widehat{D}^{1,n-2}_a$ denote the adjoint matrix of
$D^{1,n-2}_a$ (it is well known that
\[D^{1,n-2}_a\widehat{D}^{1,n-2}_a=\widehat{D}^{1,n-2}_a
D^{1,n-2}_a=\det D^{1,n-2}_a\cdot I_{n-1}=B^2_n(a)\cdot
I_{n-1}\enspace,\] where $I_{n-1}$ is the identity matrix of size
$n-1$). Given a three-diagonal structure of $D^{1,n-2}_a$, it is
easy to compute the elements of $\widehat{D}^{1,n-2}_a$ and to see
that
\[\widehat{D}^{1,n-2}_a(1,0)=r^{-1}_1 a_1\det
D^{3,n-2}_a\quad\mbox{and}\quad\widehat{D}^{1,n-2}_a(1,n-2)=r^{-1}_2
a_2 r_3 a_3\cdot\dots\cdot r_{n-2} a_{n-2}\ .\] By
(\ref{eq:DeltaPr}), (\ref{eq:Delta}) and (\ref{eq:Delta2k}), $\det
D^{3,n-2}_a=\Delta_a(3,n-2)=(B^2_{n-2}(a))^{2^{2k}}$. Also note that
$\prod_{i=2}^{n-1}r^{(-1)^{i-1}}_i=r_0
r_1^{-1}r^{-\frac{2^{nk}+1}{2^k+1}}=r_0 r_1^{-1}\delta^{-1}$. Then,
from (\ref{eq:SystRed}) we get that
\begin{equation}
 \label{eq:deg22k}
B^2_n(a)z_2=r_0 r^{-1}_1\left(a_0 a_1(B^2_{n-2}(a))^{2^{2k}}+
\delta^{-1}\prod_{i=2}^{n-1}a_i\right)z_0\enspace.
\end{equation}

Suppose that $L_a(z)$ has more than $2^{2k}$ zeros. These are also
roots of equation (\ref{eq:deg22k}) which has degree $2^{2k}$ and
this is possible only if the latter equation is identically zero.
Thus, in particular, $B_n(a)=0$ and, by Lemma~\ref{le:ZeB},
$a=\frac{v_0^{2^{2k}+1}}{(v_0+v_1)^{2^k+1}}$ for some
$v\in\mathrm{GF}(2^{nk})\setminus\mathrm{GF}(2^k)$ with ${\rm
Tr}^{nk}_k(v_0)=0$. Also, necessarily,
 \setlength{\arraycolsep}{0.14em}
\begin{eqnarray*}
0&=&a_0 a_1(B^2_{n-2}(a))^{2^{2k}}+\delta^{-1}\prod_{i=2}^{n-1}a_i\\
&\stackrel{(\ref{eq:BiV})}{=}&\frac{v_0 v_1
v_2v_3}{(v_0+v_1)(v_1+v_2)^2(v_2+v_3)}\left(\frac{\sum_{i=1}^{n-2}v_i\prod_{i=2}^{n-3}v_i}
{\prod_{i=1}^{n-3}(v_i+v_{i+1})}\right)^{2^{2k+1}}\\
&&+\frac{\delta^{-1}\prod_{i=2}^{n-1}v_i v_{i+2}}
{(v_2+v_3)\prod_{i=3}^{n-1}(v_i+v_{i+1})^2 (v_n+v_{n+1})}\\
&=&\frac{v_0 v_1 v_2 v_3\sum_{i=3}^n v_i^2 \prod_{i=4}^{n-1}v_i^2
+(v_1+v_2)^2\delta^{-1}\prod_{i=2}^{n-1}v_i
v_{i+2}}{(v_0+v_1)(v_1+v_2)^2(v_2+v_3)\prod_{i=3}^{n-1}(v_i+v_{i+1})^2}\enspace.
\end{eqnarray*}
 \setlength{\arraycolsep}{5pt}\noindent
Thus, since ${\rm Tr}^{nk}_k(v_0)=\sum_{i=0}^{n-1}v_i=0$,
 \setlength{\arraycolsep}{0.14em}
\begin{eqnarray*}
0&=&v_0 v_1 v_2 v_3\sum_{i=3}^n v_i^2
\prod_{i=4}^{n-1}v_i^2+(v_1+v_2)^2\delta^{-1}\prod_{i=2}^{n-1}v_i v_{i+2}\\
&=&v_0 v_1 v_2 v_3(v_1+v_2)^2\prod_{i=4}^{n-1}v_i^2
+(v_1+v_2)^2\delta^{-1}v_0 v_1 v_2 v_3\prod_{i=4}^{n-1}v_i^2
\end{eqnarray*}
 \setlength{\arraycolsep}{5pt}\noindent
which leads to $\delta=r^{\frac{2^{nk}+1}{2^k+1}}=1$ which
contradicts the condition imposed on $r$.
\hspace*{\fill}${\Box}$

\begin{proposition}
 \label{pr:L22kZero}
For any $a\in\mathrm{GF}(2^{nk})$, if $Y_n(a)=0$ then $L_a(z)$ has
$1$ or $2^{2k}$ zeros in $\mathrm{GF}(2^{2nk})$. Moreover,
\[{\rm Tr}_{2nk}(rav^{2^k+1})+{\rm Tr}_{nk}(v^{2^{nk}+1})=0\]
for any $v\in\mathrm{GF}(2^{2nk})$ with $L_a(v)=0$.
\end{proposition}


{\bf Proof.} The first statement directly follows from
Proposition~\ref{pr:Lle22kZero}. Let $V\neq 0$ be a zero of
$L_a(z)$. Then all $2^{2k}$ zeros are given by $\mu V$ for every
$\mu\in\mathrm{GF}(2^{2k})$. For any $v\in\mathrm{GF}(2^{2nk})$ with
$L_a(v)=0$ we have $v=\mu V$ and
 \setlength{\arraycolsep}{0.14em}
\begin{eqnarray*}
{\rm Tr}_{2nk}\left(ra(\mu V)^{2^k+1}\right)&+&{\rm Tr}_{nk}\left((\mu V)^{2^{nk}+1}\right)=\\
&=&{\rm Tr}_{nk}\left(\mu^{2^k+1}(raV^{2^k+1}+r^{2^{nk}}a^{2^{nk}}V^{2^{(n+1)k}+2^{nk}}+V^{2^{nk}+1})\right)\\
&=&{\rm Tr}_k\left(\mu^{2^k+1}{\rm Tr}_k^{nk}(raV^{2^k+1}+r^{-1}aV^{2^{(n+1)k}+2^{nk}}+V^{2^{nk}+1})\right)\\
&=&{\rm Tr}_k(\mu^{2^k+1}Q)\enspace,
\end{eqnarray*}
 \setlength{\arraycolsep}{5pt}\noindent
where $Q={\rm
Tr}_k^{nk}(raV^{2^k+1}+r^{-1}aV^{2^{(n+1)k}+2^{nk}}+V^{2^{nk}+1})$.
We show now that $Q=0$. To this end, we define
\[U(a)=\sum_{y\in\mathrm{GF}(2^{2nk})}(-1)^{{\rm Tr}_{2nk}(ra
y^{2^k+1})+{\rm Tr}_{nk}(y^{2^{nk}+1})}\] and compute
 \setlength{\arraycolsep}{0.14em}
\begin{eqnarray*}
U^2(a)&=&\sum_{x,y\in\mathrm{GF}(2^{2nk})}(-1)^{{\rm Tr}_{2nk}(ra(x^{2^k+1}+y^{2^k+1}))+{\rm Tr}_{nk}(x^{2^{nk}+1}+y^{2^{nk}+1})}\\
&=&\sum_{y,v\in\mathrm{GF}(2^{2nk})}(-1)^{{\rm Tr}_{2nk}(ra((v+y)^{2^k+1}+y^{2^k+1}))+{\rm Tr}_{nk}((v+y)^{2^{nk}+1}+y^{2^{nk}+1})}\\
&=&\sum_{y,v\in\mathrm{GF}(2^{2nk})}(-1)^{{\rm Tr}_{2nk}(ra(v^{2^k}y+vy^{2^k}+v^{2^k+1})+yv^{2^{nk}})+{\rm Tr}_{nk}(v^{2^{nk}+1})}\\
&=&\sum_{v\in\mathrm{GF}(2^{2nk})}(-1)^{{\rm
Tr}_{2nk}(rav^{2^k+1})+{\rm Tr}_{nk}(v^{2^{nk}+1})}
\sum_{y\in\mathrm{GF}(2^{2nk})}(-1)^{{\rm Tr}_{2nk}(y^{2^k}L_a(v))}\\
&=&2^{2nk}\sum_{v\in\mathrm{GF}(2^{2nk}),\,L_a(v)=0}(-1)^{{\rm Tr}_{2nk}(rav^{2^k+1})+{\rm Tr}_{nk}(v^{2^{nk}+1})}\\
&=&2^{2nk}\sum_{\mu\in\mathrm{GF}(2^{2k})}(-1)^{{\rm
Tr}_k(\mu^{2^k+1}Q)}\enspace.
\end{eqnarray*}
 \setlength{\arraycolsep}{5pt}\noindent

Suppose that $Q\neq 0$. When $\mu$ runs through
$\mathrm{GF}(2^{2k})$ then ${\rm Tr}_k(\mu^{2^k+1}Q)$ takes on the
value zero $M_0=1+(2^k+1)(2^{k-1}-1)$ times and the value one
$M_1=(2^k+1)2^{k-1}$ times. Hence, $M_0-M_1=-2^k$ that implies
\[U(b)^2=2^{2nk}(M_0-M_1)=-2^{(2n+1)k}\]
which is impossible. Thus, $Q=0$ and the proposition is proved.
\hspace*{\fill}${\Box}$

\section{Four-Valued Cross Correlation}
 \label{sec:4VDist}
In this section, we prove our main result formulated in
Corollary~\ref{co:main}. We start by considering the following
exponential sum denoted $S_0(a)$ that to some extent is determined
by the following lemma.

\begin{lemma}
 \label{le:S0}
For an odd $n>2$ and $a\in\mathrm{GF}(2^{nk})$ let $S_0(a)$ be
defined by
\[S_0(a)=\sum_{y\in\mathrm{GF}(2^{2nk})}(-1)^{{\rm Tr}_{2nk}(ay^{2^k+1})+{\rm Tr}_{nk}(y^{2^{nk}+1})}\enspace.\]
Then
\[S_0(a)=2^{nk}\sum_{v\in\mathrm{GF}(2^{nk}),\,A_a(v)=0}(-1)^{{\rm Tr}_{nk}(v)}\enspace,\]
where $A_a(x)$ is defined in (\ref{eq:A}) with
$c^{-1}=\delta+\delta^{-1}$ for $\delta$ being a primitive
$(2^k+1)^{\rm{th}}$ root of unity over $\mathrm{GF}(2)$.
\end{lemma}

{\bf Proof.} Let $\delta$ be a primitive $(2^k+1)^{\rm{th}}$ root of
unity over $\mathrm{GF}(2)$ (note that
$\delta\in\mathrm{GF}(2^{2k})\setminus\mathrm{GF}(2^k)$). Then any
element in $\mathrm{GF}(2^{2nk})$ can be written uniquely as
$y=u+\delta v$ with $u,v\in\mathrm{GF}(2^{nk})$.

Let ${\overline y}=y^{2^{nk}}$ and
$c^{-1}=\delta+\delta^{-1}\in\mathrm{GF}(2^k)$, then we obtain
 \setlength{\arraycolsep}{0.14em}
\begin{eqnarray*}
y^{2^k+1}+{\overline y}^{2^k+1}&=&(u+\delta v)^{2^k+1}+(u+\delta^{2^k} v)^{2^k+1}\\
&=&(u^{2^k}v+uv^{2^k})(\delta+\delta^{2^k})\\
&=&c^{-1}(u^{2^k}v+uv^{2^k})
\end{eqnarray*}
 \setlength{\arraycolsep}{5pt}\noindent
and further
 \setlength{\arraycolsep}{0.14em}
\begin{eqnarray*}
y^{2^{nk}+1}&=&(u+\delta v)^{2^{nk}+1}\\
&=&u^{2^{nk}+1}+u^{2^{nk}}v\delta+uv^{2^{nk}}\delta^{2^{nk}}+v^{2^{nk}+1}\\
&=&u^2+c^{-1}uv+v^2\enspace.
\end{eqnarray*}
 \setlength{\arraycolsep}{5pt}\noindent

Hence, we get
 \setlength{\arraycolsep}{0.14em}
\begin{eqnarray*}
S_0(a)&=&\sum_{y\in\mathrm{GF}(2^{2nk})}(-1)^{{\rm Tr}_{nk}(a(y^{2^k+1}+{\overline y}^{2^k+1})+y^{2^{nk}+1})}\\
&=&\sum_{u,v\in\mathrm{GF}(2^{nk})}(-1)^{{\rm Tr}_{nk}(ac^{-1}(u^{2^k}v+uv^{2^k})+u^2+c^{-1}uv+v^2)}\\
&=&\sum_{v\in\mathrm{GF}(2^{nk})}(-1)^{{\rm Tr}_{nk}(v)}
\sum_{u\in\mathrm{GF}(2^{nk})}(-1)^{{\rm Tr}_{nk}(u^{2^k}c^{-1}(a^{2^k}v^{2^{2k}}+v^{2^k}+av+c))}\\
&=&2^{nk}\sum_{v\in\mathrm{GF}(2^{nk}),\,A_a(v)=0}(-1)^{{\rm
Tr}_{nk}(v)}\enspace,
\end{eqnarray*}
 \setlength{\arraycolsep}{5pt}\noindent
where $A_a(x)=a^{2^k}x^{2^{2k}}+x^{2^k}+ax+c$ and
$c^{-1}=\delta+\delta^{-1}$. Consider equation $x^2+c^{-1}x=1$ that
has two roots $\delta$ and $\delta^{-1}$ which are elements in
$\mathrm{GF}(2^{2k})$ but not in $\mathrm{GF}(2^k)$. Letting
$x=c^{-1}y$ we get $y^2+y=c^2$ that has two solutions $c\delta$ and
$c\delta^{-1}$ which do not belong to $\mathrm{GF}(2^k)$. Thus,
${\rm Tr}_{k}(c^2)={\rm Tr}_k(c)=1$. \hspace*{\fill}${\Box}$

We can now determine $S_0(a)$ completely in the following corollary.

\begin{corollary}
 \label{cor:S0}
Under the conditions of Lemma~\ref{le:S0} the distribution of
$S_0(a)$ is given as follows:
 \setlength{\arraycolsep}{0.0em}
\[\begin{array}{rll}
-&2^{nk}&\quad\mbox{if}\quad Z_n(a)\neq 0\ ,\\
&2^{(n+1)k}&\quad\mbox{if}\quad Z_n(a)=0\ \mbox{and}\ B_n(a)\neq 0\ ,\\
-&2^{(n+2)k}&\quad\mbox{if}\quad B_n(a)=0\enspace.
\end{array}\]
 \setlength{\arraycolsep}{5pt}\noindent
\end{corollary}

{\bf Proof.} The distribution follows immediately from
Lemma~\ref{le:S0} and the results about the roots of $A_a(x)$ proved
in Section~\ref{sec:Aff}. If $Z_n(a)\neq 0$ then we use
Proposition~\ref{pr:A1Zero}, if $Z_n(a)=0$ and $B_n(a)\neq 0$ then
Proposition~\ref{pr:A2kZero} comes in handy and, finally, if
$B_n(a)=0$ then we need Propositions~\ref{pr:l22kZero} and
\ref{pr:A22kZero}.
\hspace*{\fill}${\Box}$

\begin{lemma}
 \label{le:SiSquare}
For an odd $n>2$ and $a\in\mathrm{GF}(2^{nk})$ let
$r=\alpha^{(2^{nk}-1)2^{k-1}}$, where $\alpha$ is a primitive
element of $\mathrm{GF}(2^{2nk})$. Let also
 \setlength{\arraycolsep}{0.14em}
\begin{eqnarray*}
S_j(a)&=&\sum_{y\in\mathrm{GF}(2^{2nk})}(-1)^{{\rm Tr}_{2nk}(r^j
ay^{2^k+1}) +{\rm Tr}_{nk}(y^{2^{nk}+1})}\quad\mbox{and}\\
S_{2^k+1-j}(a)&=&\sum_{y\in\mathrm{GF}(2^{2nk})}(-1)^{{\rm
Tr}_{2nk}(r^{-j}ay^{2^k+1}) +{\rm Tr}_{nk}(y^{2^{nk}+1})}
\end{eqnarray*}
 \setlength{\arraycolsep}{5pt}\noindent
for $j=1,2,\dots,2^{k-1}$. Then
\renewcommand{\theenumi}{\roman{enumi}}
\renewcommand{\labelenumi}{(\theenumi)}
\begin{enumerate}
\item\label{it:4} $S_j(a)=S_{2^k+1-j}(a)\quad$ for
$j\in\{1,\dots,2^{k-1}\}$ and
\item\label{it:5} $S_i(a)^2=2^{2nk} T_a\enspace,$ \medskip\newline where $T_a$ is the number of zeros in
$\mathrm{GF}(2^{2nk})$ of $L_a(z)$ defined in (\ref{eq:L}) with
$r^i$ (resp. $r^{-(2^k+1-i)}$) taken for $r$ if $1\leq i\leq
2^{k-1}$ (resp. $2^{k-1}<i\leq 2^k$).
\end{enumerate}
\end{lemma}

{\bf Proof.} (i) For any $j\in\{1,\dots,2^{k-1}\}$, straightforward
calculations give
 \setlength{\arraycolsep}{0.14em}
\begin{eqnarray*}
S_j(a)&=&\sum_{y\in\mathrm{GF}(2^{2nk})}(-1)^{{\rm Tr}_{2nk}(r^{j2^{nk}}a^{2^{nk}}y^{(2^k+1)2^{nk}})
+{\rm Tr}_{nk}(y^{(2^{nk}+1)2^{nk}})}\\
&=&\sum_{x\in\mathrm{GF}(2^{2nk})}(-1)^{{\rm Tr}_{2nk}(r^{-j}ax^{2^k+1})+{\rm Tr}_{nk}(x^{2^{nk}+1})}\\
&=&S_{2^k+1-j}(a)\enspace.
\end{eqnarray*}
 \setlength{\arraycolsep}{5pt}\noindent

(ii) For any $i\in\{1,\dots,2^k\}$ let $d(i)=i$ if
$i\in\{1,\dots,2^{k-1}\}$ and $d(i)=-(2^k+1-i)$ if
$i\in\{2^{k-1}+1,\dots,2^k\}$. Exactly the same way as in the
calculations of $U^2(a)$ in Proposition~\ref{pr:L22kZero} we obtain
 \setlength{\arraycolsep}{0.14em}
\begin{eqnarray*}
S_i(a)^2&=& \sum_{x,y\in\mathrm{GF}(2^{2nk})}
(-1)^{{\rm Tr}_{2nk}(r^{d(i)}a(x^{2^k+1}+y^{2^k+1}))+{\rm Tr}_{nk}(x^{2^{nk}+1}+y^{2^{nk}+1})}\\
&=&2^{2nk}\sum_{v\in\mathrm{GF}(2^{2nk}),\,L_a(v)=0}(-1)^{{\rm
Tr}_{2nk}(r^{d(i)}av^{2^k+1})+{\rm Tr}_{nk}(v^{2^{nk}+1})}\enspace,
\end{eqnarray*}
 \setlength{\arraycolsep}{5pt}\noindent
where
$L_a(z)=z^{2^{(n+1)k}}+r^{d(i)2^k}a^{2^k}z^{2^{2k}}+r^{d(i)}az$.
And, finally, we have
\[{\rm
Tr}_{2nk}(r^{d(i)}av^{2^k+1})+{\rm Tr}_{nk}(v^{2^{nk}+1})=0\] for
any zero $v\in\mathrm{GF}(2^{2nk})$ of $L_a(z)$, by
Propositions~\ref{pr:L1Zero} and \ref{pr:L22kZero}.
\hspace*{\fill}${\Box}$

We are now in position to completely determine the distribution of
$S(a)$ defined in (\ref{eq:S}) for $a\in\mathrm{GF}(2^{nk})^*$.
Since this is equivalent to the distribution of $C_d(\tau)+1$ for
$\tau=0,1,\dots,2^{nk}-2$, our main result in
Corollary~\ref{co:main} is a consequence of the theorem below.

\begin{theorem}
 \label{th:S(a)main}
Let $m=2nk$ and $d=\frac{2^{nk}+1}{2^k+1}$, where $n>2$ is odd and
$k>1$. Then the exponential sum $S(a)$ defined in (\ref{eq:S}) for
$a\in\mathrm{GF}(2^{nk})^*$ (and $C_d(\tau)+1$, for
$\tau=0,1,\dots,2^{nk}-2$) have the following distribution:
 \setlength{\arraycolsep}{0.0em}
\[\begin{array}{rllll}
-&2^{(n+1)k}&\quad\mbox{occurs}\quad&\frac{2^{(n-1)k}-1}{2^{2k}-1}&\quad\mbox{times}\ ,\\
-&2^{nk}&\quad\mbox{occurs}\quad&\frac{(2^{nk}-1)(2^{k-1}-1)}{2^k-1}&\quad\mbox{times}\ ,\\
&0&\quad\mbox{occurs}\quad&2^{(n-1)k}-1&\quad\mbox{times}\ ,\\
&2^{nk}&\quad\mbox{occurs}\quad&\frac{(2^{nk}+1)2^{k-1}}{2^k+1}&\quad\mbox{times}\enspace.
\end{array}\]
 \setlength{\arraycolsep}{5pt}\noindent
\end{theorem}

{\bf Proof.} Take $\alpha$ being a primitive element of
$\mathrm{GF}(2^{2nk})$ and let ${\rm ind}(x)$ be defined as
$x=\alpha^{{\rm ind}(x)}$ for any $x\in\mathrm{GF}(2^{2nk})$.
Letting also $r=\alpha^{(2^{nk}-1)2^{k-1}}$ we observe that
$r^{2^{nk}+1}=1$ and that $r^i$ is not a $(2^k+1)^{\rm{th}}$ power
in $\mathrm{GF}(2^{2nk})$ for any $i=1,2,\dots,2^{k-1}$ since ${\rm
ind}(r^i)\equiv i\ (\bmod\;2^k+1)$. It is also clear that ${\rm
ind}(r^{-i})\equiv 2^{2nk}-1-{\rm ind}(r^i)\equiv 2^k+1-i\
(\bmod\;2^k+1)$.

Finding the distribution of the cross-correlation function
$C_d(\tau)+1$ is equivalent to computing the distribution of $S(a)$
defined in (\ref{eq:S}) for $a\in\mathrm{GF}(2^k)^*$. To calculate
$S(a)$, we first observe that $\gcd(2^k+1,2^m-1)=2^k+1$. If we first
let $x=y^{2^k+1}$ then $x=r^i y^{2^k+1}$ and, finally,
$x=r^{-i}y^{2^k+1}$ for $i=1,\ldots,2^{k-1}$ and $y$ running through
$\mathrm{GF}(2^m)$ then $x$ will run through $\mathrm{GF}(2^m)$ in
total $2^k+1$ times. Further, since $d(2^k+1)(2^{nk}+1)\equiv
2(2^{nk}+1)\ (\bmod\;2^m-1)$, we obtain
 \setlength{\arraycolsep}{0.14em}
\begin{eqnarray*}
(2^k+1)S(a)&=&\sum_{i=0}^{2^{k-1}}\sum_{y\in\mathrm{GF}(2^m)}(-1)^{{\rm
Tr}_m(r^i ay^{2^k+1})+{\rm Tr}_{nk}(y^{2^{nk}+1})}\\
&+&\sum_{i=1}^{2^{k-1}}\sum_{y\in\mathrm{GF}(2^m)}(-1)^{{\rm
Tr}_m(r^{-i}ay^{2^k+1})+{\rm Tr}_{nk}(y^{2^{nk}+1})}
=\sum_{i=0}^{2^k}S_i(a)\enspace,
\end{eqnarray*}
 \setlength{\arraycolsep}{5pt}\noindent
where $S_i(a)$ are defined as in Lemma~\ref{le:SiSquare}. We divide
the proof into three cases.

{\bf Case 1:} ($B_n(a)=0$)

In this case, Corollary~\ref{cor:S0} gives $S_0(a)=-2^{(n+2)k}$ and,
by Proposition~\ref{pr:L1Zero}, $L_a(z)$ has exactly one zero (since
$Z_n(a)=0$ by (\ref{eq:A(B)})). Therefore, by
Lemma~\ref{le:SiSquare} (\ref{it:5}), $S_i(a)=\pm 2^{nk}$ for all
values of $i=1,2,\dots,2^k$. Thus,
\[(2^k+1)S(a)=-2^{(n+2)k}+t2^{nk}\enspace,\]
where $|t|\leq 2^k$. Reduce both sides of the latter identity modulo
$2^k+1$ to obtain $1-t\equiv 0\ (\bmod\;2^k+1)$. Since $t$ is even
then $t\neq 1$ and the only possibility is $t=-2^k$ leading to
$S(a)=-2^{(n+1)k}$.

{\bf Case 2:} ($Z_n(a)=0$ and $B_n(a)\neq 0$)

In this case, Corollary~\ref{cor:S0} gives $S_0(a)=2^{(n+1)k}$ and,
by Proposition~\ref{pr:L1Zero}, $L_a(z)$ has exactly one zero.
Therefore, by Lemma~\ref{le:SiSquare} (\ref{it:5}), $S_i(a)=\pm
2^{nk}$ for all values of $i=1,2,\dots,2^k$. Thus,
\[(2^k+1)S(a)=2^{(n+1)k}+t2^{nk}\enspace,\]
where $|t|\leq 2^k$. Reduce both sides of the latter identity modulo
$2^k+1$ to obtain $1-t\equiv 0\ (\bmod\;2^k+1)$. Since $t$ is even
then $t\neq 1$ and the only possibility is $t=-2^k$ leading to
$S(a)=0$.

{\bf Case 3:} ($Z_n(a)\neq 0$)

In this case, Corollary~\ref{cor:S0} gives $S_0(a)=-2^{nk}$.
Consider a set of $2^{k-1}$ values
\[Y_n(a)=Z^2_n(a)+{\rm N}^{nk}_k(a)(\delta^j+\delta^{-j})\quad
\mbox{for}\quad j=1,2,\dots,2^{k-1}\enspace,\] where
$\delta=\alpha^{\frac{(2^{2nk}-1)2^{k-1}}{2^k+1}}$ is an element of
multiplicative order $2^k+1$. If quadratic equation
$x+x^{-1}=Z^2_n(a)/{\rm N}^{nk}_k(a)$ has two solutions then the
product of these is one and thus, there is at most one zero (say,
when $j=\mathcal{J}$) in this set. Therefore, by
Propositions~\ref{pr:L1Zero}, \ref{pr:L22kZero} and
Lemma~\ref{le:SiSquare} (\ref{it:5}), $S_i(a)=\pm 2^{nk}$ for all
values of $i=1,2,\dots,2^k$ except for, possibly, two with
$i=\mathcal{J}$ and $i=2^k+1-\mathcal{J}$ when
$S_\mathcal{J}(a)=S_{2^k+1-\mathcal{J}}(a)=\pm 2^{(n+1)k}$, using
Lemma~\ref{le:SiSquare} (\ref{it:4}).

In the case when $S_i(a)=\pm 2^{nk}$ for all values of
$i=1,2,\dots,2^k$, we have
\[(2^k+1)S(a)=-2^{nk}+t2^{nk}\enspace,\]
where $|t|\leq 2^k$. Reduce both sides of the latter identity modulo
$2^k+1$ to obtain $1-t\equiv 0\ (\bmod\;2^k+1)$. Since $t$ is even
then $t\neq 1$ and the only possibility is $t=-2^k$ leading to
$S(a)=-2^{nk}$.

Finally, in the case when $S_i(a)=\pm 2^{nk}$ for all values of
$i=1,2,\dots,2^k$ except for two, we have
\[(2^k+1)S(a)=-2^{nk}+t2^{nk}+\varepsilon 2^{(n+1)k+1}\enspace,\]
where $|t|\leq 2^k-2$ and $\varepsilon\in\{-1,+1\}$. Reduce both
sides of the latter identity modulo $2^k+1$ to obtain
$1-t+2\varepsilon\equiv 0\ (\bmod\;2^k+1)$. Since $t$ is even then
$t\notin\{-1,3\}$ and the only possibility is $t=-(2^k-2)$ and
$\varepsilon=1$ leading to $S(a)=2^{nk}$.

The three cases above give, in total, the possible values $0$, $\pm
2^{nk}$ and $-2^{(n+1)k}$ for $S(a)$. Suppose the cross-correlation
function $C_d(\tau)+1$ takes on the value zero $r$ times, the value
$2^{nk}$ is taken on $s$ times, the value $-2^{nk}$ occurs $t$ times
and the value $-2^{(n+1)k}$ occurs $v$ times. Since
$S(a)=-2^{(n+1)k}$ is possible only in Case 1, when $B_n(a)=0$,
then, by Lemma~\ref{le:ZeB}, $v=\frac{2^{(n-1)k}-1}{2^{2k}-1}$. By
Proposition~\ref{pr:A2kZero}, the number of
$a\in\mathrm{GF}(2^{nk})^*$ such that $Z_n(a)=0$ and $B_n(a)\neq 0$
is equal $2^{(n-1)k}-1$. Thus, since $S(a)=0$ is possible only in
Case 2, when $Z_n(a)=0$ and $B_n(a)\neq 0$, then $r=2^{(n-1)k}-1$.

For the remaining values of $S(a)=\pm 2^{nk}$, obviously,
\[s+t=2^{nk}-1-(r+v)=\frac{2^{(n+2)k}-2^{(n+1)k}-2^{nk}+1}{2^{2k}-1}\enspace.\]
On the other hand, from Lemma~\ref{le:pow1} it follows that
\[2^{nk}s-2^{nk}t-2^{(n+1)k}v=2^{nk}(s-t)-\frac{2^{(n+1)k}(2^{(n-1)k}-1)}{2^{2k}-1}=2^{nk}\]
and
\[s-t=1+\frac{2^k(2^{(n-1)k}-1)}{2^{2k}-1}=\frac{2^{nk}+2^{2k}-2^k-1}{2^{2k}-1}\enspace.\]
Thus, the solutions are $s=\frac{(2^{nk}+1)2^{k-1}}{2^k+1}$ and
$t=\frac{(2^{nk}-1)(2^{k-1}-1)}{2^k-1}$. \hspace*{\fill}${\Box}$

The arguments in this paper also work for $k=1$. However, in this
case, the corresponding decimation $d=(2^n+1)/3$ is only
three-valued (see \cite{NeHe06}). Indeed, in this case,
\[3S(a)=S_0(a)+S_1(a)+S_2(a)=S_0(a)+2S_1(a)\enspace.\]
It was proved in Proposition~\ref{pr:A1Zero} that
$Z_n(a)\in\mathrm{GF}(2^k)=\mathrm{GF}(2)$. Thus, if $Z_n(a)\neq 0$
then $Z_n(a)=1$ and the value of $S_1(a)$ is defined by whether
\[Y_n(a)=Z^2_n(a)+{\rm N}^{n}_1(a)(\delta+\delta^{-1})=1+\delta+\delta^{-1}\enspace,\]
where $\delta=\alpha^{(2^{2n}-1)/3}$ is primitive in
$\mathrm{GF}(4)$, is zero or not (note that $a\neq 0$). Since
$\delta+1=\delta^2=\delta^{-1}$, we have $\delta+\delta^{-1}=1$.
Therefore, if $Z_n(a)\neq 0$ then $S(a)=2^n$ and is never equal to
$-2^n$. This reduces the four-valued cross-correlation case to three
values.

\section{Conclusions}
We have identified new pairs of $m$-sequences having different
lengths $2^{2nk}-1$ and $2^{2k}-1$, where $n>2$ is odd and $k>1$,
with four-valued cross correlation and we have completely determined
the cross-correlation distribution. These pairs differ from the
sequences in the Kasami family by the property that instead of the
decimation $d=1$ we take $d=\frac{2^{nk}+1}{2^k+1}$.


\begin{thebibliography}{10}
\providecommand{\url}[1]{#1} \csname url@samestyle\endcsname
\providecommand{\newblock}{\relax} \providecommand{\bibinfo}[2]{#2}
\providecommand{\BIBentrySTDinterwordspacing}{\spaceskip=0pt\relax}
\providecommand{\BIBentryALTinterwordstretchfactor}{4}
\providecommand{\BIBentryALTinterwordspacing}{\spaceskip=\fontdimen2\font
plus \BIBentryALTinterwordstretchfactor\fontdimen3\font minus
  \fontdimen4\font\relax}
\providecommand{\BIBforeignlanguage}[2]{{%
\expandafter\ifx\csname l@#1\endcsname\relax
\typeout{** WARNING: IEEEtran.bst: No hyphenation pattern has been}%
\typeout{** loaded for the language `#1'. Using the pattern for}%
\typeout{** the default language instead.}%
\else \language=\csname l@#1\endcsname \fi #2}}
\providecommand{\BIBdecl}{\relax} \BIBdecl

\bibitem{He76} T.~Helleseth, ``Some results about the
    cross-correlation function between two
  maximal linear sequences,'' \emph{Discrete Mathematics}, vol.~16, no.~3, pp.
  209--232, Nov. 1976.

\bibitem{HeKu98} T.~Helleseth and P.~V. Kumar, ``Sequences with low
    correlation,'' in
  \emph{Handbook in Coding Theory}, V.~S. Pless and W.~C. Huffman, Eds.\hskip
  1em plus 0.5em minus 0.4em\relax Amsterdam: Elsevier Science B.V., 1998,
  vol.~II, ch.~21, pp. 1765--1853.

\bibitem{DoFeHeRo06} H.~Dobbertin, P.~Felke, T.~Helleseth, and
    P.~Rosendahl, ``Niho type
  cross-correlation functions via {D}ickson polynomials and {K}loosterman
  sums,'' \emph{{IEEE} Trans. Inf. Theory}, vol.~52, no.~2, pp. 613--627, Feb.
  2006.

\bibitem{NeHe06} G.~J. Ness and T.~Helleseth, ``Cross correlation of
    $m$-sequences of different
  lengths,'' \emph{{IEEE} Trans. Inf. Theory}, vol.~52, no.~4, pp. 1637--1648,
  Apr. 2006.

\bibitem{Ka66} T.~Kasami, ``Weight distribution formula for some
    classes of cyclic codes,''
  Coordinated Science Laboratory, University of Illinois, Urbana, Tech. Rep.
  R-285 (AD 637524), Apr. 1966.

\bibitem{NeHe06_1} G.~J. Ness and T.~Helleseth, ``A new three-valued
    cross correlation between
  $m$-sequences of different lengths,'' \emph{{IEEE} Trans. Inf. Theory},
  vol.~52, no.~10, pp. 4695--4701, Oct. 2006.

\bibitem{HeKhNe07} T.~Helleseth, A.~Kholosha, and G.~J. Ness,
    ``Characterization of $m$-sequences
  of lengths $2^{2k}-1$ and $2^k-1$ with three-valued crosscorrelation,''
  \emph{{IEEE} Trans. Inf. Theory}, vol.~53, no.~6, pp. 2236--2245, Jun. 2007.

\bibitem{NeHe07} G.~J. Ness and T.~Helleseth, ``A new family of
    four-valued cross correlation
  between $m$-sequences of different lengths,'' \emph{{IEEE} Trans. Inf.
  Theory}, vol.~53, no.~11, pp. 4308--4313, Nov. 2007.

\bibitem{IlKu85} V.~P. Il'in and Y.~I. Kuznetsov,
    \emph{Three-Diagonal Matrices and their
  Applications}.\hskip 1em plus 0.5em minus 0.4em\relax Moscow: Nauka, 1985,
  (in Russian).

\bibitem{Bl04} A.~W. Bluher, ``On $x^{q+1}+ax+b$,'' \emph{Finite
    Fields and Their
  Applications}, vol.~10, no.~3, pp. 285--305, Jul. 2004.

\bibitem{LiNi97} R.~Lidl and H.~Niederreiter, \emph{Finite Fields},
    ser. Encyclopedia of
  Mathematics and its Applications.\hskip 1em plus 0.5em minus 0.4em\relax
  Cambridge: Cambridge University Press, 1997, vol.~20.

\end{thebibliography}


\end{document}